\documentclass[preprint,one-column,nofootinbib,english]{revtex4-1}
\usepackage{amsthm,amsmath,amsfonts,dsfont,upgreek} 
\usepackage{amssymb,epsfig,setspace}
\usepackage{graphicx}
\usepackage{babel}
\usepackage{verbatim}
\usepackage{mathrsfs}
\usepackage{enumitem}
\usepackage{bm}
\usepackage{color}
\usepackage{xcolor}

\begin{document}

\hyphenation{ano-ther ge-ne-ra-te dif-fe-rent know-le-d-ge po-ly-no-mi-al}
\hyphenation{me-di-um or-tho-go-nal as-su-ming pri-mi-ti-ve pe-ri-o-di-ci-ty}
\hyphenation{mul-ti-p-le-sca-t-te-ri-ng i-te-ra-ti-ng e-q-ua-ti-on}
\hyphenation{wa-ves di-men-si-o-nal ge-ne-ral the-o-ry sca-t-te-ri-ng}
\hyphenation{di-f-fe-r-ent tra-je-c-to-ries e-le-c-tro-ma-g-ne-tic pho-to-nic}
\hyphenation{Ray-le-i-gh di-n-ger Kra-jew-ska Wal-czak Ham-bur-ger Ad-di-ti-o-nal-ly}
\hyphenation{Kon-ver-genz-the-o-rie ori-gi-nal in-vi-si-b-le cha-rac-te-ri-zed}
\hyphenation{sa-ti-s-fy}
\hyphenation{Ne-ver-the-less sa-tu-ra-te E-ner-gy}

\title{Constraint polynomial approach - an alternative to the functional Bethe Ansatz method?}

\author{Alexander Moroz}

\affiliation{Wave-scattering.com}

\author{Andrey E. Miroshnichenko}

\affiliation{School of Engineering and Information Technology,
University of New South Wales Canberra
Northcott Drive, Campbell, ACT 2600, Australia
}
 
\begin{abstract}
Recently developed general constraint polynomial approach is shown to replace
a set of algebraic equations of the functional Bethe Ansatz method by a single polynomial constraint.
As the proof of principle, the usefulness of the method is demonstrated 
for a number of quasi-exactly solvable (QES) potentials of the Schr\"odinger equation, 
such as two different sets of modified Manning potentials with three parameters, 
an electron in Coulomb and magnetic fields and relative motion of two electrons in an external oscillator potential,
the hyperbolic Razavy potential, and a (perturbed) double sinh-Gordon system.
The approach enables one to straightforwardly determine eigenvalues and wave functions.
Odd parity solutions for the modified Manning potentials are also determined. 
For the QES examples considered here, constraint polynomials terminate a 
finite chain of orthogonal polynomials in an independent variable that need not to be necessarily energy. 
In the majority of cases the finite chain of orthogonal polynomials is characterized by 
a positive-definite moment functional ${\cal L}$, implying that a corresponding
constraint polynomial has only real and simple zeros. Constraint polynomials are 
shown to be different from the weak orthogonal Bender-Dunne polynomials.
At the same time the QES examples considered elucidate essential difference
with various generalizations of the Rabi model. Whereas in the former case there are $n+1$ 
polynomial solutions at each point of a $n$th baseline, in the latter case there are at most 
$n+1$ polynomial solutions on entire $n$th baseline.
\end{abstract}


%

\maketitle

\section{Introduction}
\label{sc:intr}
The Schr\"odinger equation ($\hbar=2m=1$)
\begin{equation}
\left(-\frac{d^2}{dx^2}+V\right)\psi=E\psi
\label{SE}
\end{equation}
for a number of quasi-exactly solvable potentials $V$ 
can on using a suitable substitution be recast in the same basic form as 
\cite{HaS,PBss,BP,BPdw,KK,Ishk18,IshcH,Ishk17}
\begin{equation}
(a_3z^3 +a_2z^2+a_1z) \frac{d^2\phi(z)}{dz^2}+(b_2z^2+b_1z+b_0) 
 \frac{d\phi(z)}{dz}+(c_1z+c_0) \phi(z) = 0, 
\label{qese}
\end{equation}
where $a_3, a_2, a_1, b_2, b_1, b_0, c_1, c_0$ are constant parameters.
This form corresponds to the general Heun equation \cite{T16,TPR16,Ishk18}, and its confluent \cite{IshcH}
and bi-confluent \cite{Ishk17} forms, provided that one of the regular singular
points is at $z=0$. Eq. (\ref{qese}) is a particular type 
of more general ordinary differential equation (ODE) with polynomial
coefficients for which a general concept of gradation slicing has been recently 
employed in order to analyze their polynomial solutions \cite{AMcp}.
Gradation slicing is universal and easily applicable algorithmic
recursive approach for obtaining polynomial solutions which does not require 
any a priori knowledge about hidden 
algebraic structure of ODE. Its usefulness has been so far demonstrated 
on the examples of various Rabi models \cite{AMcp}. 

In the present article we employ the gradation slicing approach of Ref. \cite{AMcp}
to determine polynomial solutions of quasi-exactly solvable (QES)
Schr\"odinger equation for Xie \cite{Xi} and Chen et al. \cite{CWX} 
three parameters modified Manning potentials \cite{Mnn2,HaS,BP},
an electron in Coulomb and magnetic fields and relative 
motion of two electrons in an external oscillator potential \cite{Tr94,CH},
the perturbed double sinh-Gordon system (DSHG) \cite{HaS,BP,KM},
and the hyperbolic Razavy potential \cite{HaS,BP,Rzv}. All those QES potentials lend themselves 
to $sl_2$ algebraization \cite{T16,TPR16}.
At the same time the above QES examples are used to elucidate essential difference
with various generalizations of the Rabi model \cite{AMcp}. Whereas in the former case there are $n+1$ 
polynomial solutions at each point of a $n$th baseline [defined by condition (\ref{bnc}) below], 
in the latter case there are at most $n+1$ polynomial solutions on entire $n$th baseline. 
In both cases a given baseline characterizes the set of model parameter 
in which case an $sl_2$ algebraization with a given spin is possible.
The difference between QES examples and Rabi models arises due to a cardinally 
different qualitative behaviour 
under variations of a spectral parameter. (The latter can be either energy, as 
in examples of Sec. \ref{sc:wop}, or some other model parameter, 
as in the so-called {\em coupling constant metamorphosis} examples of Sec. \ref{sc:nwop}.)
When corresponding spectral parameter is varied then, in the QES examples presented here,
one remains at a {\em fixed} point of a baseline. In other words, a corresponding algebraic Heun operator 
remains unchanged. Contrary to that, for a number of generalizations of the Rabi model \cite{AMcp}
variations of spectral parameter induce translation on the corresponding baseline.
This has the effect that to different values of spectral parameter there correspond {\em different}
algebraic Heun operators (cf. Sec. \ref{sec:dissl2}). 
It is deemed expedient to appreciate this difference as it has led
to occasional confusion in published literature.

Another motivation behind the present article is to provide an alternative 
to the functional Bethe Ansatz method 
\cite{Xi,CWX,Tr94,CH,HaS,BP,KM}. Indeed, the eigenvalues, eigenfunctions and 
the allowed potential 
parameters were previously given exclusively in terms of the roots of a set 
of algebraic Bethe Ansatz equations of the functional Bethe Ansatz method 
\cite{Xi,CWX,Tr94,CH,HaS,BP,KM}. It is demonstrated here that the set of algebraic Bethe Ansatz equations 
can be efficiently replaced by a recurrence [cf. Eq. (\ref{gm0rcp}) below] together with 
a single polynomial constraint ${\cal P}=0$ 
[cf. Eq. (\ref{spc}) below]. In general solving for the roots of ${\cal P}(n)=0$ determines 
an isolated finite set of points in parameter space at which polynomial solutions are possible.

In what follows, we first recapitulate the gradation slicing approach of Ref. \cite{AMcp}
in Sec. \ref{sc:gsa}. Then the approach is illustrated on the coupling 
constant metamorphosis QES examples in Sec. \ref{sc:nwop}
and QES examples with energy as spectral parameter in Sec. \ref{sc:wop}. 
Some important issues are discussed in Sec. \ref{sec:disc}. 
In particular, a relation to $sl_2$ algebraization and an algebraic Heun operator is discussed in
Sec. \ref{sec:dissl2}. A discussion of when ${\cal P}(n)$ has necessarily 
only {\em real} and {\em simple} roots can be found in Sec. \ref{sec:discsz}. A comparison of ${\cal P}(n)$ and
the so called weak orthogonal polynomials of Lancosz-Haydock and Bender-Dunne is 
provided in Sec. \ref{sec:discwo}. We then conclude with Sec. \ref{sec:conc}.
For the sake or presentation, a number of intermediary calculations has been 
relegated to online supplementary material.

\section{Summary of gradation slicing approach}
\label{sc:gsa}
General necessary and sufficient conditions for the existence of a polynomial solution 
have been recently formulated involving constraint relations \cite{AMcp}. 
In the terminology of Ref. \cite{AMcp}, the {\em grade} of a term $z^m d_z^l$ is integer $m-l$.
One can straightforwardly identify that the respective terms 
of the differential operator ${\cal L}$ on the left-hand side of Eq. (\ref{qese}) 
have the highest grade $\gamma=1$, the lowest grade $\gamma_*=-1$, and 
can be assembled into three slices with the grades $1,0,-1$ with the respective multiplicators
\begin{eqnarray}
& F_1(n)=n(n-1)a_3+nb_2+c_1, \qquad F_0(n)=n(n-1)a_2+nb_1 +c_0, &
\nonumber\\
& F_{-1}(n)=n(n-1)a_1+nb_0. & 
\label{grmchp}
\end{eqnarray}

In general, the necessary conditions for the ODE (\ref{qese}) with the grade $\gamma=1$ to have a 
polynomial solution is that for some $n\in\mathbb{N}$
\begin{equation}
F_1(n)=0.
\label{bnc}
\end{equation}
Solving the condition (\ref{bnc}) usually imposes a constraint on model parameters,
which may include energy \cite{Jd,AMcp}.
The condition $F_1(n)=0$ is known as the {\em baseline} condition 
for the Rabi models \cite{AMcp,Ks} and for Jahn-Teller systems \cite{Jd}, 
because it constraints allowable energies to a set of lines, or hyperplanes, in a parameter space. 
The necessary {\em baseline} condition reappears also in the 
functional Bethe Ansatz method (cf. Theorem 4 and Remark 9 of Ref. \cite{AMcp}; 
Eqs. (1.8-10) of Ref. \cite{Zh}), or as one of the conditions of $sl_2$ algebraization \cite{T16,AMcp,Zh6}
[cf. Sec. \ref{sec:dissl2} for more details].

The necessary conditions for the ODE (\ref{qese}) with the grade $\gamma=1$ to have a {\em unique} 
polynomial solution of degree $n\ge 1$ is that (cf. Theorems 1 and 2 of \cite{AMcp}),
\begin{equation}
F_1(n)=0,\qquad F_1(k)\ne 0,\qquad 0\le k<n.
\label{th1nsc}
\end{equation}
The conditions enable one to determine 
unique set of coefficients $\{P_{nk}\}_{k=0}^n$, defined recursively by the {\em three-term} recurrence 
relations (TTRR) for $1\le k\le n$, beginning with $P_{n0}=1$ (cf. Eq. (11) of Ref. \cite{AMcp})
\begin{align}
P_{n1} & =-F_0(n)P_{n0}/F_1(n-1),
\nonumber\\
 P_{n2} & =- [ F_{-1}(n)P_{n0}+ F_0 (n-1) P_{n1} ]/F_1(n-2),
\nonumber\\
\qquad & \vdots\qquad \qquad \vdots \qquad \qquad \vdots\qquad \qquad \vdots 
\nonumber\\
 P_{n,k} & =-[F_{-1}(n+2-k) P_{n,k-2}+F_0 (n+1-k) P_{n,k-1}]/F_1(n-k),
\nonumber\\
\qquad & \vdots\qquad \qquad \vdots \qquad \qquad \vdots\qquad \qquad \vdots 
\nonumber\\
 P_{nn}& =-[F_{-1}(2) P_{n,n-2}+F_0(1) P_{n,n-1}]/F_1 (0).
\label{gm0rcp}
\end{align}
If the unique (monic) polynomial solution exists, then it is necessarily given by 
(cf. Theorems 1 and 2 of \cite{AMcp})
\begin{equation}
S_n(z)=\prod_{i=1}^n(z-z_i)=\sum_{k=0}^n P_{n,n-k} z^k \qquad \qquad (P_{n0}\equiv 1).
\label{snex}
\end{equation}
The parameters entering the recurrence coefficients $F_{\mathfrak g}(k)$ in (\ref{gm0rcp}) are 
assumed to satisfy the $F_1(n)=0$ constraint. 

The conditions (\ref{th1nsc}) become both necessary and {\em sufficient} conditions for the ODE (\ref{qese})
to have a unique polynomial solution, provided that some subset of model parameters satisfying (\ref{bnc}) 
obeys additionally (cf. Eq. (16) of Ref. \cite{AMcp})
\begin{equation}
{\cal P}(n):= F_{-1}(1) P_{n,n-1} +F_0(0) P_{nn}=b_0 P_{n,n-1} +c_0 P_{nn}=0.
\label{spc}
\end{equation}
This equation can be seen as continuation of the TTRR (\ref{gm0rcp}) one step further by 
formally defining $P_{n,n+1}=-{\cal P}(n)$.

The coefficients $F_{\mathfrak g}(k)$ are {\em polynomials} in model parameters
[e.g. examples (\ref{3mF}), (\ref{3mFo}), (\ref{3mFv}), (\ref{3CF}), (\ref{3mFvo}),
(\ref{RzvF}), (\ref{DSFpm}), (\ref{KMFp}) below]. Hence ${\cal P}(n)$ multiplied by 
$\prod_{k=n-1}^0 F_\gamma(k)\ne 0$ is necessarily a polynomial in model parameters, too.
For the examples considered here it will be shown that the coefficients 
$F_{\mathfrak g}(k)$ of Eq. (\ref{grmchp}) confined to a given baseline generate by the TTRR (\ref{gm0rcp}) 
a finite orthogonal polynomial system $\{P_{nk},\, k=0,1,2,\ldots,n,\, {\cal P}(n)\}$ 
in some spectral parameter. The spectral parameter is a model parameter
that does {\em not} enter the constraint $F_1(n)=0$, and in fact none of multiplicators $F_1(k)$.
Hence a multiplication of ${\cal P}(n)$ by $\prod_{k=n-1}^0 F_\gamma(k)\ne 0$
is in fact not necessary, because ${\cal P}(n)$ is already a polynomial in the spectral parameter. 

For the models considered here we have the following {\em dichotomy}:
\begin{itemize}

\item[({\bf A1})] $F_1(n)$ does depend on energy. Hence by solving the constraint $F_1(n)=0$
energy can be expressed as a function of model parameters, $E=E(V_j)$, and thereby 
eliminated from recurrence (\ref{gm0rcp}) and from the constraint polynomial (\ref{spc}). 
In these examples $E$ is {\em not} spectral parameter and we have
the above mentioned coupling constant metamorphosis.
It turns out that the corresponding spectral parameter is a model parameter 
that does {\em not} enter the constraint $F_1(n)=0$. 
(For example, in the Manning potential case of Sec. \ref{sc:3pM} (i) one fixes $V_1$ and $V_2$ 
together with energy $E(V_1,V_2)$ and (ii) searches for the roots of 
the constraint polynomial (\ref{spc}) as a function of $V_3$ - cf. Figs. \ref{fg1}, \ref{fg2}.) 

\item[({\bf A2})] If only the multiplicator $F_0(k)$ depends on energy, and is a {\em linear} function 
of it, then $E$ {\em is} the spectral parameter.

\end{itemize}
An important characteristics of all examples considered here is that as 
the spectral parameter varies one stays at a {\em fixed} point of an $n$th baseline.
Constraint polynomial ${\cal P}(n)$ will be shown to terminate
a finite orthogonal polynomial system in corresponding spectral parameter.
In the case of alternative ({\bf A1}), and is some examples of alternative ({\bf A2}),
${\cal P}(n)$ will be 
shown to have only {\em real} and {\em simple} roots.
The constraint polynomial relation (\ref{spc}) then
determines a {\em discrete} set of $n+1$th spectral parameter values 
at which polynomial solutions exist
at any given {\em fixed} point of the $n$th baseline.
Thereby a set of algebraic Bethe Ansatz equations can be replaced by 
a single polynomial constraint (\ref{spc}).

The constraint relation (\ref{spc}) in the case of alternative ({\bf A2}) 
provides a kind of quantization rule for the energy levels.
The latter sounds similar to the role played by a critical polynomial 
of the Lanczos-Haydock finite-chain of polynomials \cite{Hd,Lnz} (more known as the
Bender-Dunne polynomials \cite{BD,AMhd}). Yet, as discussed in Sec. \ref{sec:discwo}, 
such a resemblance is only coincidental.

\section{Examples of $F_1(n)$ depending on energy resulting in a
coupling constant metamorphosis}
\label{sc:nwop}

\subsection{A modified Manning potential with three parameters}
\label{sc:3pM}
In this section we examine parity invariant potential
\begin{equation}
V(x) = -V_1 \mbox{ sech}^6 x-V_2 \mbox{ sech}^4 x-V_3 \mbox{ sech}^2 x
\label{XmMp} 
\end{equation}
studied by Xie \cite{Xi}, which for $V_1=0$ reduces to the Manning potential \cite{Mnn2}.
Obviously $\lim_{|x|\to\infty} V(x)=0$.
This potential describes a double-well potential whenever
$V_1>0$, $V_2<0$, $V_3>0$ and $-V_3/(2V_2)<1$. The two minima of the potential
are then located at $x_\pm=\pm\mbox{arcsech }\sqrt{-V_3/(2V_2)}$.
\begin{figure}
\begin{center}
\includegraphics[width=14cm,clip=0,angle=0]{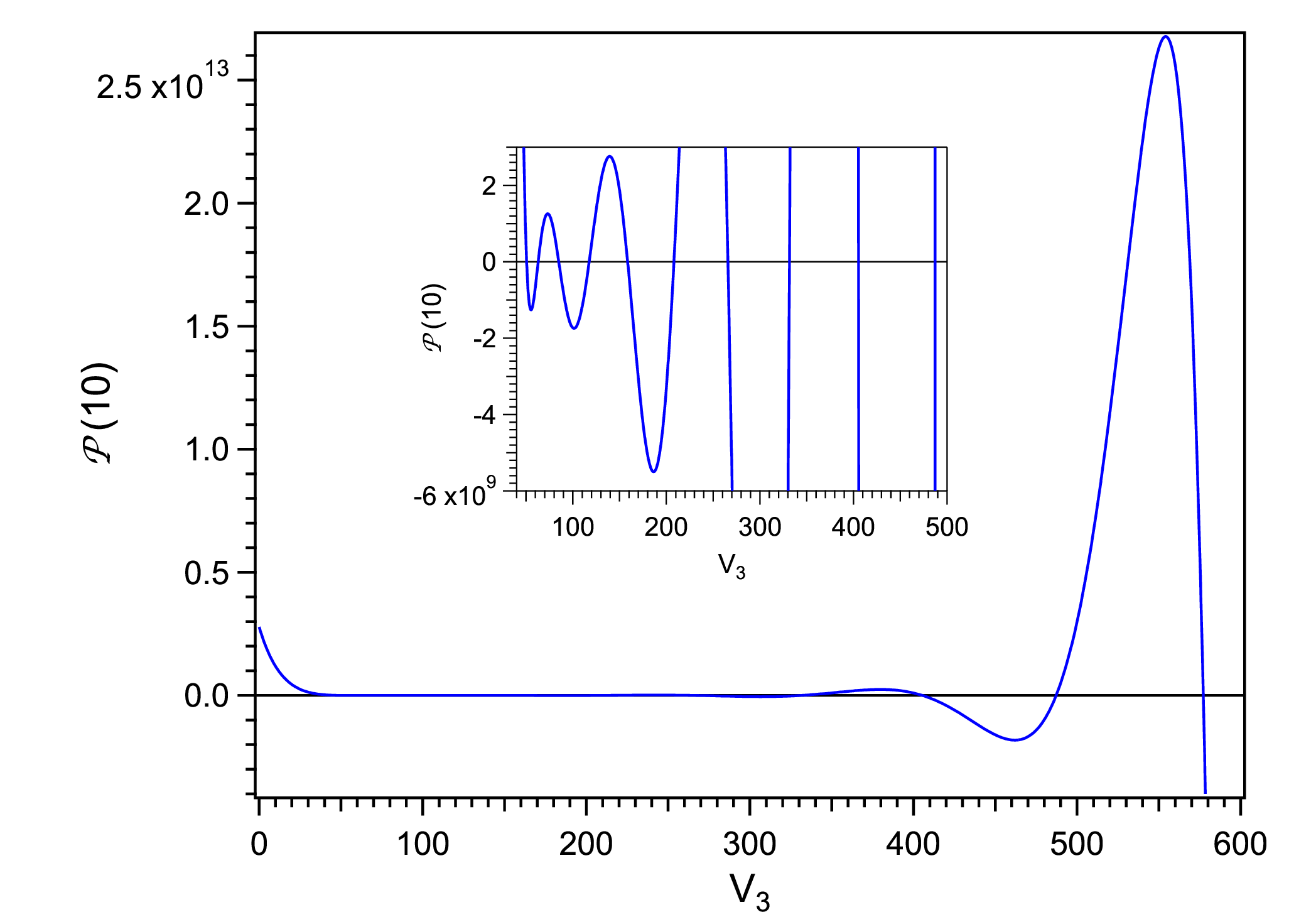}
\end{center}
\caption{Constraint polynomial for the Xie generalized Manning potential in the {\em even} parity case 
as a function of $V_3$ with fixed $V_1=1$, $V_2=-50$, and $n=10$. There are $11$ {\em real} roots for
$V_3=50.6499,\, 62.9912,\, 85.016,\, 117.499,\, 158.65,\, 208.126,\, 265.78,\, 331.54,\, 
405.368,\, 487.239,\, 577.141$. They all correspond to the real eigenvalue 
$\sqrt{-E_{10}}=3$ [cf. Eq. (\ref{epMe})]. For the real roots we have 
$V_1>0$, $V_2<0$, $V_3>0$. The double-well condition is satisfied for the lowest three values of $V_3$. 
Corresponding wave functions are shown in Fig. \ref{fgXMwf}.
}
\label{fg1}
\end{figure}

\subsubsection{Even parity solutions}
\label{sc:3pMe}
The substitution 
\begin{equation}
\psi(x) = \exp\left( \frac{\sqrt{V_1}}{2}\, \tanh^2 x\right) (1-\tanh^2 x)^{\frac{\sqrt{-E}}{2}} \phi(x)
\label{XiA}
\end{equation}
followed by the change in variable through $z=\tanh^2 x$
transform the Schr\"odinger equation (\ref{SE}) into (\ref{qese}) with \cite{Xi}
\begin{eqnarray}
& a_2 = 4,\qquad a_1 = -4,& 
\nonumber\\
& b_2 = 4\sqrt{V_1},\qquad b_1 = 6+4(\sqrt{-E}-\sqrt{V_1}),\qquad b_0 = -2,& 
\nonumber\\
& c_1 = V_1+V_2+3\sqrt{V_1}+2\sqrt{V_1}\sqrt{-E},\quad 
c_0 = \sqrt{-E} -E -\sqrt{V_1} -V_1 -V_2-V_3.& 
\label{XMec}
\end{eqnarray}
In the Ansatz (\ref{XiA}) and further below the principal branch of fractional powers is assumed.

\begin{figure}
\begin{center}
\includegraphics[width=14cm,clip=0,angle=0]{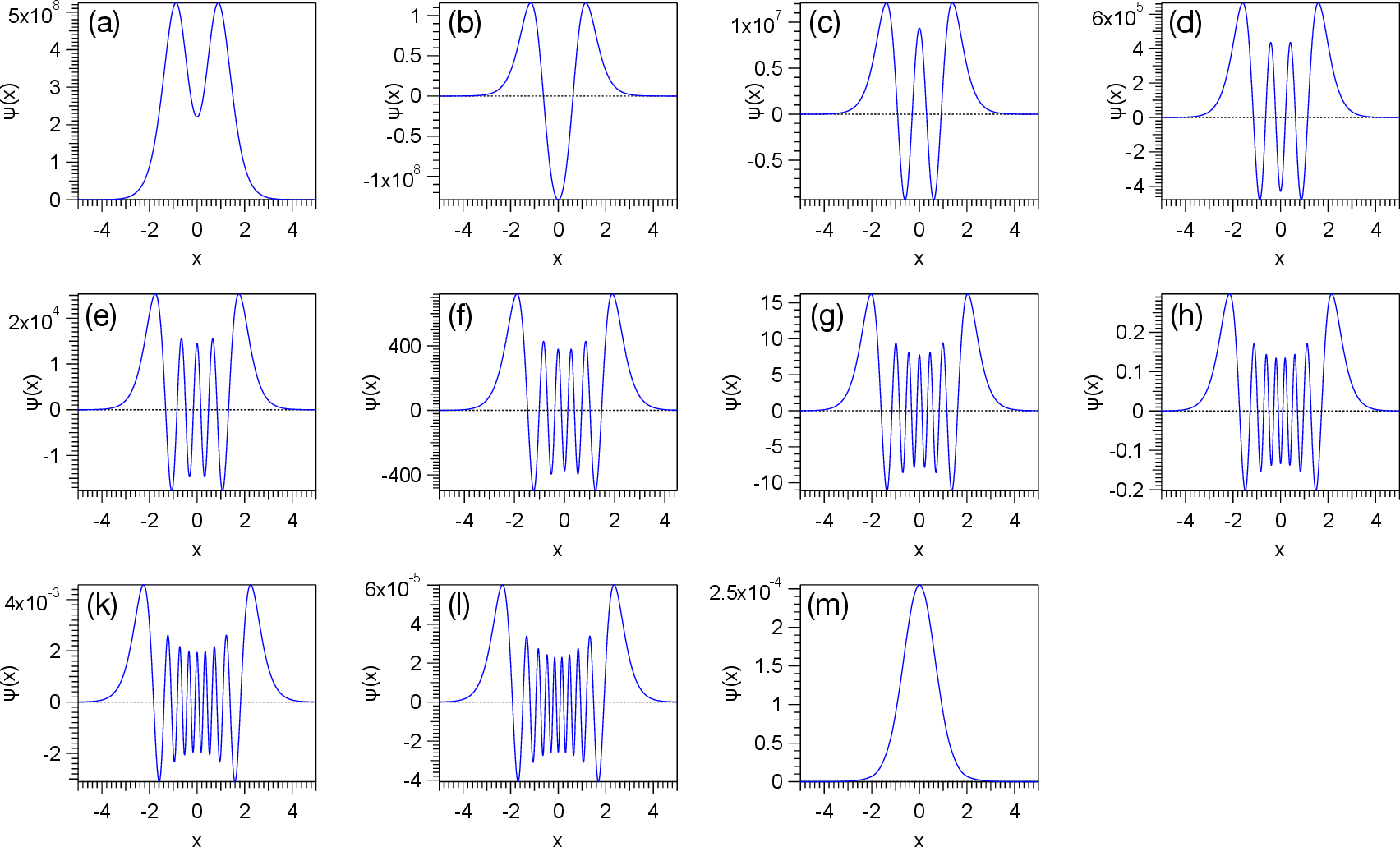}
\end{center}
\caption{Even parity polynomial eigenfunctions for the Xie generalized Manning potential 
with fixed $V_1=1$, $V_2=-50$, and $n=10$ for the values of 
$V_3=50.6499,\, 62.9912,\, 85.016,\, 117.499,\, 158.65,\, 208.126,\, 265.78,\, 331.54,\, 
405.368,\, 487.239,\, 577.141$ as in Fig. \ref{fg1}.
They all correspond to the real eigenvalue $\sqrt{-E_{10}}=3$ [cf. Eq. (\ref{epMe})].
The double-well condition is satisfied for the lowest three values of $V_3$.
}
\label{fgXMwf}
\end{figure}
Because $c_1$ is energy dependent, the necessary condition (\ref{bnc}),
\begin{equation}
F_1(n)=4n\sqrt{V_1}+V_1+V_2+3\sqrt{V_1}+2\sqrt{V_1}\sqrt{-E} =0,
\nonumber
\end{equation}
forces energy onto a $n$th baseline, 
\begin{equation}
\sqrt{-E_n}=-2n-\frac{V_1+V_2+3\sqrt{V_1}}{2\sqrt{V_1}}-\frac32 \longrightarrow
-2(n+1)-\frac{V_2}{2} \qquad (V_1\rightarrow 1).
\label{epMe}
\end{equation}
Because $\lim_{|x|\to\infty} \tanh^2 x=1$ and $1-\tanh^2 x=\cosh^{-2} x$, 
the solutions expressed by the Ansatz (\ref{XiA}) are normalizable
for any polynomial $\phi(x)$ as long as $\sqrt{-E}>0$. With a fixed value of $V_1>0$,
the normalizability condition requires
\begin{equation}
V_2 < -\left[(4n+3)\sqrt{V_1}+ V_1\right].
\label{xmnce}
\end{equation}

On the $n$th baseline one has in virtue of (\ref{grmchp})
\begin{eqnarray}
& F_1(k)=-4(n-k) \sqrt{V_1}, \qquad F_0(k)=2k [2k+1+2(\sqrt{-E_n}-\sqrt{V_1})] 
+c_0(n), &
\nonumber\\
& F_{-1}(k)=-2k(2k-1),& 
\label{3mF}
\end{eqnarray}
where, given $\sqrt{-E} -E =\sqrt{-E} (\sqrt{-E} +1)$,
\begin{equation}
c_0(n)=\left(2n+\frac{V_1+V_2}{2\sqrt{V_1}}+\frac32 \right) 
\left(2n +\frac{V_1+V_2}{2\sqrt{V_1}}+\frac12\right) -\sqrt{V_1}
 -V_1-V_2-V_3.
\label{c0eM}
\end{equation}
Being a linear function, $F_1(k)$ has for each $n$ only single zero. Hence the conditions (\ref{th1nsc})
are satisfied and there is always a unique polynomial solution for a given fixed set of parameters.

Given the above expression for $c_0$, an obvious choice of independent variable, or spectral parameter, is $V_3$.
The choice of $V_3$ immediately implies that one remains at a fixed point of the baseline,
because neither the baseline nor resulting energy does not depend on the value of $V_3$. 
The choice of any of $\sqrt{V_1}$ and $V_2$ as independent variable would be analogous to what happens in search
of the exceptional spectrum of the Rabi model \cite{AMcp,Ks,Jd}. This option is discussed
later in Sec. \ref{sec:dissl2}.

It turned out straightforward to reproduce the even parity roots $V_3$ of 
the constraint polynomial in Tab. 1 of \cite{BP} 
for $n=0$, $V_1=1$, $V_2=-6$, $n=1$, $V_1=1$, $V_2=-12$, and $n=2$, $V_1=1$, $V_2=-18$.
It took not much effort to produce results of Fig. \ref{fg1} showing the constraint polynomial 
as a function of $V_3$ for fixed $V_1=1$, $V_2=-50$, and $n=10$. Fig. \ref{fgXMwf} shows wave functions
corresponding to the roots of the constraint polynomial of Fig. \ref{fg1}.

\subsubsection{Odd parity solutions}
\label{sc:3pMo}
Given that the odd parity solution has to have only odd powers of $\tanh x$, replacing 
$\phi(x)$ in the Ansatz (\ref{XiA}) by $\tanh x\, \phi(x)$ leads to a grade $\gamma=1$ and width $w=3$
differential operator for the {\em odd} parity solutions,
\begin{eqnarray}
\lefteqn{
4z(z-1)d_z^2 +\left\{z\left[4z\sqrt{V_1}+4(\sqrt{-E}-\sqrt{V_1}) +10\right] -6 \right\}d_z
}
\nonumber\\
&&
\, + z\left[V_1+V_2 + \sqrt{V_1}(5 +2\sqrt{-E}) \right]-E+3\sqrt{-E}+2-(V_1+V_2+V_3+3\sqrt{V_1}),
\nonumber\\
&&
\end{eqnarray}
where $d_z=d/dz$. The Schr\"odinger equation (\ref{SE}) is again transformed into (\ref{qese}) with
\begin{eqnarray}
& a_2 = 4,\qquad a_1 = -4,& 
\nonumber\\
& b_2 = 4\sqrt{V_1},\qquad b_1 = 10+4(\sqrt{-E}-\sqrt{V_1}),\qquad b_0 = -6,& 
\nonumber\\
& c_1 = V_1+V_2+5\sqrt{V_1}+2\sqrt{V_1}\sqrt{-E}, & 
\nonumber\\
& c_0 = -E + 3\sqrt{-E}+2 -3\sqrt{V_1} -V_1 -V_2-V_3.& 
\label{XMoc}
\end{eqnarray}
Because $c_1$ is energy dependent, the necessary condition (\ref{bnc}),
\begin{equation}
F_1(n)=4n\sqrt{V_1}+V_1+V_2+5\sqrt{V_1}+2\sqrt{V_1}\sqrt{-E} =0,
\nonumber
\end{equation}
forces energy onto a $n$th baseline, 
\begin{equation}
\sqrt{-E_n}=-2n-\frac{V_1+V_2+5\sqrt{V_1}}{2\sqrt{V_1}}\longrightarrow
-2n-3-\frac{V_2}{2} \qquad (V_1\rightarrow 1).
\label{opMe}
\end{equation}
With a fixed value of $V_1>0$, the normalizability condition requires
[cf. (\ref{xmnce})]
\begin{equation}
V_2 < -\left[(4n+5)\sqrt{V_1}+ V_1\right].
\nonumber 
\end{equation}
\begin{figure}
\begin{center}
\includegraphics[width=14cm,clip=0,angle=0]{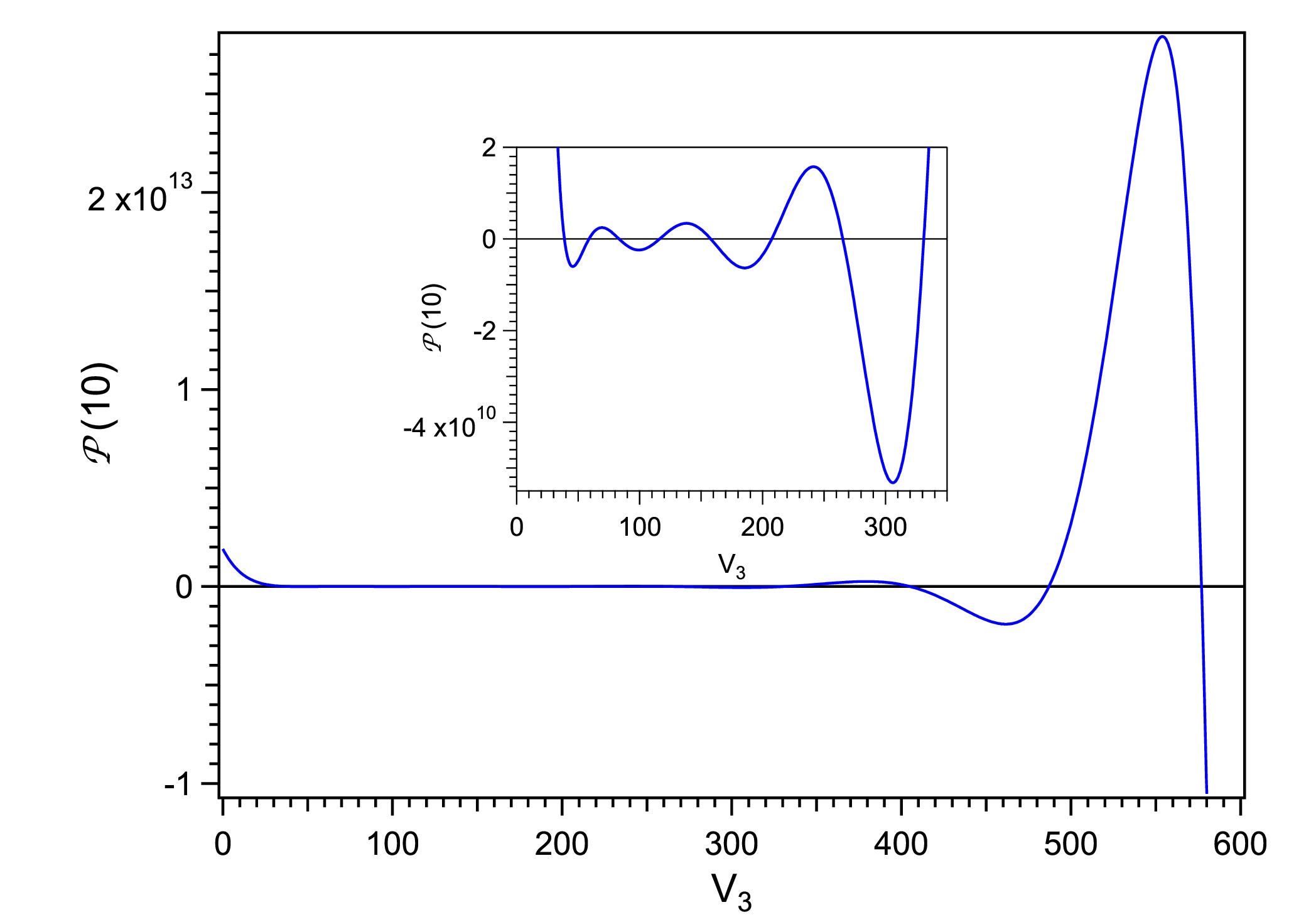}
\end{center}
\caption{Constraint polynomial for the Xie generalized Manning potential in the {\em odd} parity case
as a function of $V_3$ with fixed $V_1=1$, $V_2=-50$, and $n=10$. There are $11$ {\em real} roots for
$V_3=38.8277,\, 58.8256,\, 83.2712,\, 116.335,\, 157.819,\, 207.504,\, 265.299,\, 331.158,\, 405.056,\, 486.981,\, 576.924$. 
They all correspond to the real eigenvalue 
$\sqrt{-E_{10}}=2$ [cf. Eq. (\ref{opMe})].
The double-well condition is satisfied for the lowest three values of $V_3$. 
Corresponding wave functions are shown in Fig. \ref{fgXMwfo}.
}
\label{fg2}
\end{figure}
\begin{figure}
\begin{center}
\includegraphics[width=14cm,clip=0,angle=0]{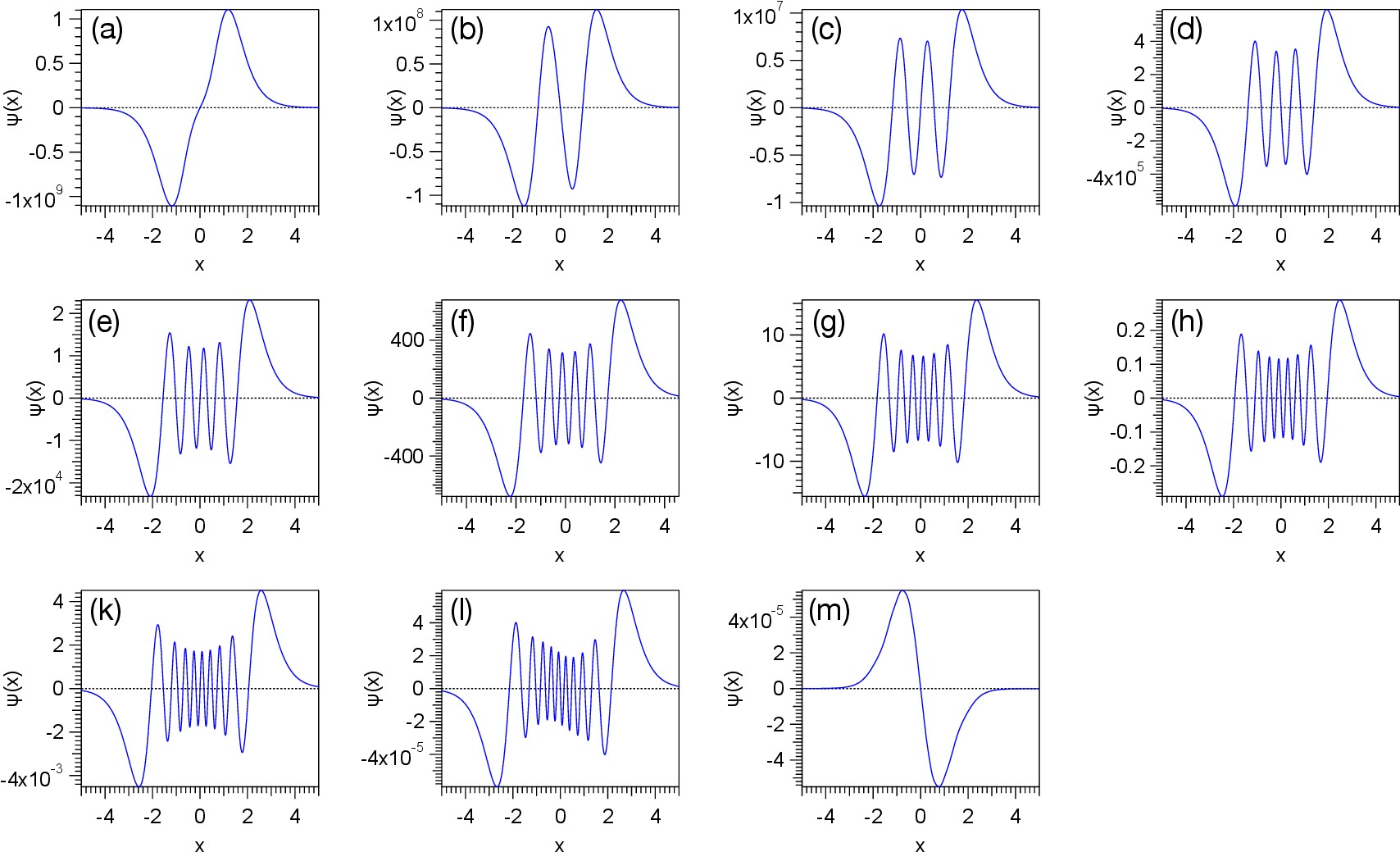}
\end{center}
\caption{Polynomial eigenfunctions for the Xie generalized Manning potential 
 in the {\em odd} parity case with fixed $V_1=1$, $V_2=-50$, and $n=10$ for the eleven values of 
$V_3=38.8277,\, 58.8256,\, 83.2712,\, 116.335,\, 157.819,\, 207.504,\, 265.299,\, 331.158,\, 405.056,\, 486.981,\, 576.924$
as in Fig. \ref{fg2}. They all correspond to the real eigenvalue $\sqrt{-E_{10}}=2$ [cf. Eq. (\ref{opMe})].
The double-well condition is satisfied for the lowest three values of $V_3$.
}
\label{fgXMwfo}
\end{figure}
On the $n$th baseline one has in virtue of (\ref{grmchp})
\begin{eqnarray}
& F_1(k)=-4(n-k) \sqrt{V_1}, \qquad F_0(k)=2k [2k+3+2(\sqrt{-E_n}-\sqrt{V_1})] 
+c_0(n), &
\nonumber\\
& F_{-1}(k)=-2k(2k+1),& 
\label{3mFo}
\end{eqnarray}
where, given $3\sqrt{-E} -E +2=(\sqrt{-E}+2) (\sqrt{-E} +1)$,
\begin{equation}
c_0(n)=\left(2n+\frac{V_1+V_2}{2\sqrt{V_1}}+\frac32 \right) 
\left(2n +\frac{V_1+V_2}{2\sqrt{V_1}}+\frac12\right) -3\sqrt{V_1} -V_1 -V_2-V_3.
\label{c0oM}
\end{equation}
Again, any solution expressed by such an amended Ansatz will be normalizable
for any polynomial $\phi(x)$ whenever $\sqrt{-E}>0$.
Fig. \ref{fg2} shows the constraint polynomial as a function of $V_3$
for fixed $V_1=1$, $V_2=-50$, and $n=10$. Fig. \ref{fgXMwfo} shows wave functions
corresponding to the roots of the constraint polynomial of Fig. \ref{fg2}.

\subsection{Chen et al. modified Manning potential with three parameters}
\label{sc:3pMa}
In this section we examine parity invariant potential
\begin{equation}
V(x) = \frac{V_1}{\cosh^2 x }+ \frac{V_2}{1+g \cosh^2 x }+ \frac{V_3}{(1+g\cosh^2 x)^2}
\label{CmMp} 
\end{equation}
studied by Chen et al. \cite{CWX}, which approximates the Manning potential \cite{Mnn2} 
in the limit $g\gg 1$. As in the previous case, $\lim_{|x|\to\infty} V(x)=0$.

\subsubsection{Even parity solutions}
\label{sc:3pCMe}
The change in variable through $z=-\sinh^2 x$ and the substitution \cite{CWX}
\begin{eqnarray}
& \psi(x) = (\cosh^2 x)^{\lambda_1} (1+g\cosh^2 x)^{\lambda_2} \phi(z),& 
\label{CMpA}\\
& \lambda_1=\frac14\left(1+\sqrt{1-4V_1}\right)\ge \frac14,\qquad 
 \lambda_2=\frac12\left[1-\sqrt{1+V_3/(1+g)}\right],& 
\label{ld12}
\end{eqnarray}
transform the Schr\"odinger equation (\ref{SE}) into (\ref{qese}) with \cite{CWX}
\begin{eqnarray}
& a_3 = 1,\qquad a_2 = -2-1/g,\qquad a_1 = 1+1/g,& 
\nonumber\\
& b_2 = 2\lambda_1+2\lambda_2+1,&
\nonumber\\
&
b_1 =- 1-\frac{1}{2g}-\left(2\lambda_1+\frac12 \right) \left(1+\frac{1}{g}\right)-2\lambda_2
 =-\left(2\lambda_1+2\lambda_2+\frac32+\frac{2\lambda_1+1}{g}\right),& 
\nonumber\\
& b_0 = \frac{1+g}{2g},\qquad c_1 = (\lambda_1+\lambda_2)^2+\frac{E}{4},&
\nonumber\\
& 
c_0 = -\frac{1+g}{4g}
 \left[2\lambda_1+\frac{2\lambda_2 g-V_2}{1+g}-V_1-\frac{V_3}{(1+g)^2} +E\right].& 
\label{CmMc}
\end{eqnarray}
The Ansatz (\ref{CMpA}) provides a normalizable solution on the interval $x\in (-\infty,\infty)$
for a polynomial $\phi(z)$ of $n$-th degree if and only if $\lambda_1+\lambda_2+n<0$.

Because $c_1$ is energy dependent, the necessary condition (\ref{bnc}),
\begin{equation}
F_1(n)=n(n-1)+ n(2\lambda_1+2\lambda_2+1)+(\lambda_1+\lambda_2)^2+\frac{E}{4} =0,
\nonumber
\end{equation}
forces energy onto a $n$th baseline,
\begin{equation}
E_n=- 4[n^2+ 2n(\lambda_1+\lambda_2)+(\lambda_1+\lambda_2)^2]=- 4(n+\lambda_1+\lambda_2)^2.
\label{eCmMe}
\end{equation}
On the $n$th baseline one has in virtue of (\ref{grmchp}) for $0\le k+k_g <n$
\begin{eqnarray}
& F_1(k)=k^2-n^2 +2(k-n)(\lambda_1+\lambda_2)>(n-k)^2>0, &
\nonumber\\
&
F_0(k)=-k(k-1) \left(2+\frac{1}{g}\right)-k\left(2\lambda_2+2\lambda_1+\frac32+\frac{2\lambda_1+1}{g}\right) 
+c_0(n), &
\nonumber\\
& F_{-1}(k)=k(k-1) \left(1+\frac{1}{g}\right) + \frac12\, k \left(1+\frac{1}{g}\right)=\frac{1+g}{2g}\, k(2k-1),& 
\label{3mFv}
\end{eqnarray}
where $k_g=1,0,-1$ denotes the subscript of corresponding $F_{k_g}$, and
\begin{equation}
c_0(n)=-\frac{1+g}{4g}
 \left[2\lambda_1+\frac{2\lambda_2 g-V_2}{1+g}-V_1-\frac{V_3}{(1+g)^2}
- 4(n+\lambda_1+\lambda_2)^2 \right].
\nonumber
\end{equation}
$F_1(k)$ is quadratic function of $k$ which has only single nonnegative root $k=n$ [the other is 
$k=-n-2(\lambda_1+\lambda_2)<0$]. Because $F_1(k)$ has for each $n$ only single nonnegative zero, 
the conditions (\ref{th1nsc}) are satisfied and there can always be only a unique polynomial solution.
\begin{figure}
\begin{center}
\includegraphics[width=14cm,clip=0,angle=0]{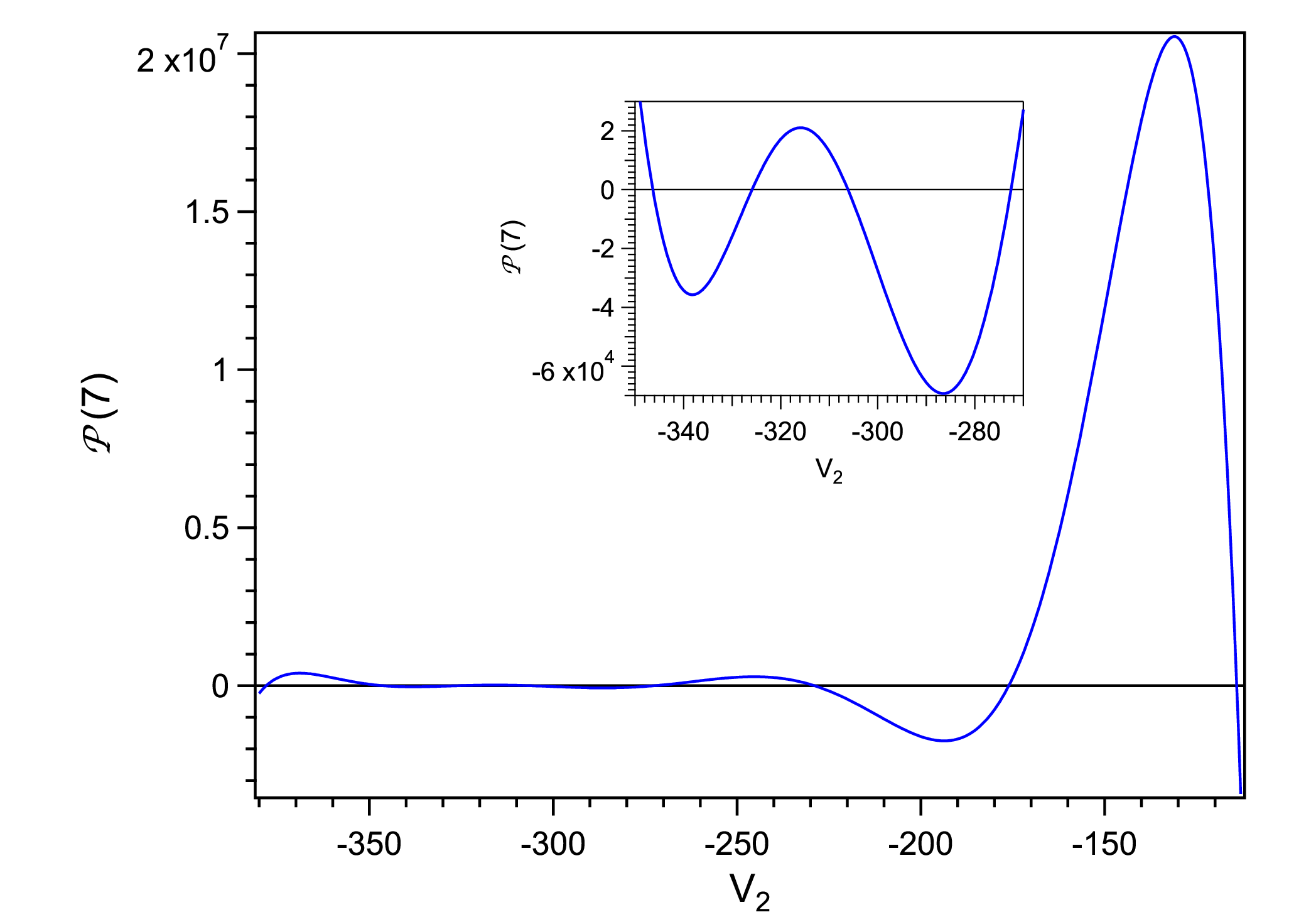}
\end{center}
\caption{Constraint polynomial for the Chen et al. generalized Manning potential 
in the {\em even} parity case as a function of $V_2$ with fixed $V_1=0.09$, $V_3=400$, $g=0.25$, and $n=7$. 
There is the maximum number of $8$ {\em real} zeros of the constraint polynomial:
$V_2=- 378.075,\, - 346.334,\, - 325.892,\, - 306.113,\, - 272.536,\, - 228.953,\, - 176.075,\, - 114.078$.
Corresponding polynomial eigenfunctions are shown in Fig. \ref{fgCMwfe}.
}
\label{fg3}
\end{figure}
\begin{figure}
\begin{center}
\includegraphics[width=14cm,clip=0,angle=0]{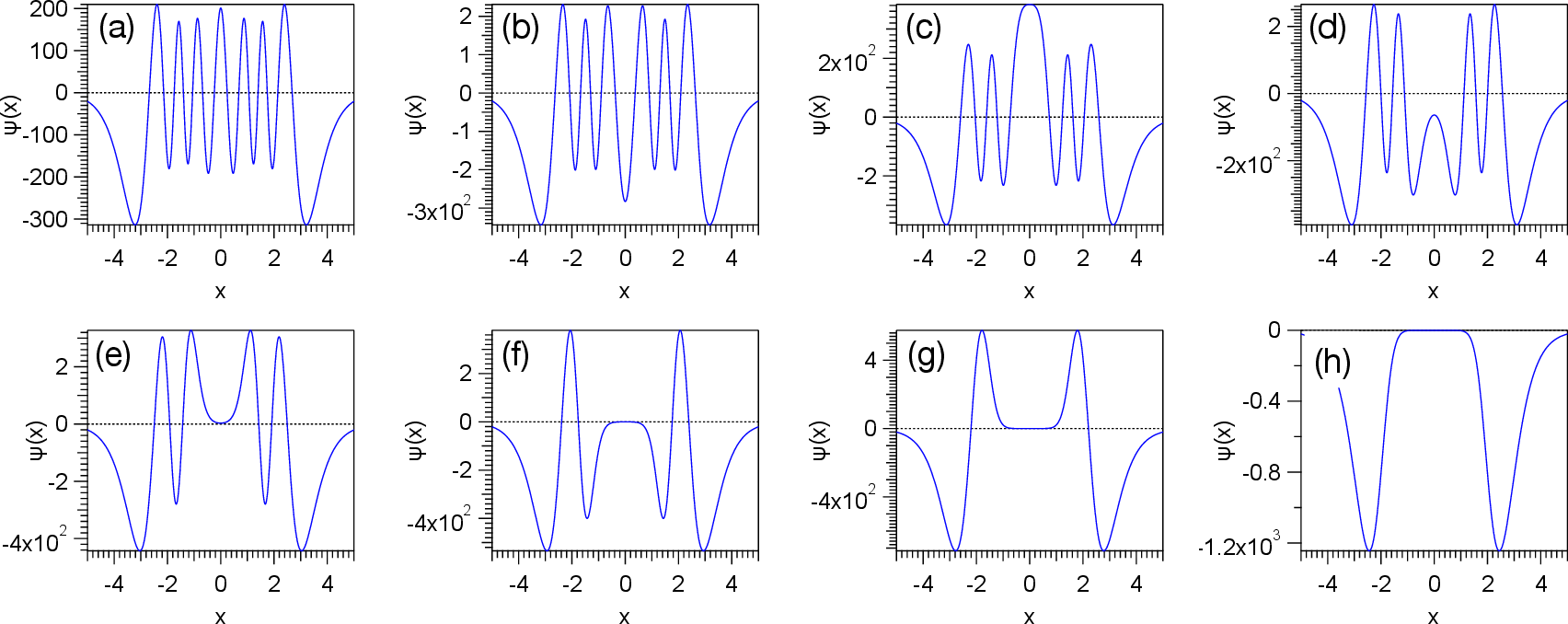}
\end{center}
\caption{Polynomial eigenfunctions for the Chen et al. generalized Manning potential 
in the {\em even} parity case with fixed $V_1=1$, $V_3=400$, $g=0.25$, and $n=7$ for the eight values of 
$V_2=- 378.075,\, - 346.334,\, - 325.892,\, - 306.113,\, - 272.536,\, - 228.953,\, - 176.075,\, - 114.078$
as in Fig. \ref{fg3}.
}
\label{fgCMwfe}
\end{figure}

Given the definition (\ref{ld12}) of $\lambda_1$ it is obvious that one has to have $V_1\le 1/4$ in order
that $\lambda_1\in\mathbb{R}$. The latter restriction has been satisfied by all the cases (I to III) considered 
by Chen et al. \cite{CWX}.

It turned out straightforward to reproduce the even parity roots $V_2$ of the constraint 
polynomial in Tab. 1 of \cite{BPdw} for $V_1=0.09$, $V_3=10$, $g=0.25$ and $n=0,1,2,3$. 
Fig. \ref{fg3} shows the constraint polynomial as a function of $V_2$
for fixed $V_1=1$, $V_3=400$, $g=0.25$, and $n=7$. Fig. \ref{fgCMwfe} displays wave functions
corresponding to the roots of the constraint polynomial of Fig. \ref{fg3}.

\subsubsection{Odd parity solutions}
\label{sc:3pCMo}
Obviously the Ansatz (\ref{CMpA}) can lead to only {\em even} parity solutions.
In order to arrive at {\em odd} parity solutions it is, given $z=-\sinh^2 x$, 
expedient to modify the Ansatz by adding an extra $\sinh x$ factor,
\begin{eqnarray}
& \psi(x) = (\cosh x)^{2\lambda_1} (1+g\cosh^2 x)^{\lambda_2} \sinh x\, \phi(z),& 
\label{CMpAe}
\end{eqnarray}
with $\lambda_1$ and $\lambda_2$ as in (\ref{CMpA}).
The Ansatz (\ref{CMpAe}) yields a normalizable solution on the interval $x\in (-\infty,\infty)$ 
for a polynomial $\phi(z)$ of $n$-th degree if and only if $\lambda_1+\lambda_2+n<-1/2$.

According to (\ref{DtB}) and (\ref{DtC})
\begin{eqnarray}
\Delta B(z) &=& z^2 -z\, \frac{1+2g}{g}+\frac{1+g}{g},
\nonumber\\
\Delta C(z) &=& z \left(\lambda_1+\lambda_2+\frac14\right) 
 - \left(\lambda_1+\lambda_2+\frac14\right)\frac{1+g}{g}+ \frac{\lambda_2}{g}\cdot
\nonumber
\end{eqnarray}
Therefore in the expressions in (\ref{CmMc}) the coefficients $a_j$ remain the same,
whereas the $b_j$ and $c_j$ coefficients are amended to
\begin{eqnarray}
& b_2 = 2(\lambda_1+\lambda_2+1),&
\nonumber\\
&
b_1 =-\left[ 2\lambda_1+2\lambda_2+\frac72+\frac{2(\lambda_1+1)}{g}\right],& 
\nonumber\\
& b_0 = \frac{3(1+g)}{2g},\qquad c_1 = (\lambda_1+\lambda_2)(\lambda_1+\lambda_2+1) + \frac{E+1}{4},&
\nonumber\\
& 
c_0 = -\frac{1+g}{4g}
 \left[6\lambda_1+4\lambda_2+1+\frac{2\lambda_2 g-V_2}{1+g}-V_1-\frac{V_3}{(1+g)^2} +E\right]
+\frac{\lambda_2}{g}\cdot & 
\label{CmMco}
\end{eqnarray}
\begin{figure}
\begin{center}
\includegraphics[width=14cm,clip=0,angle=0]{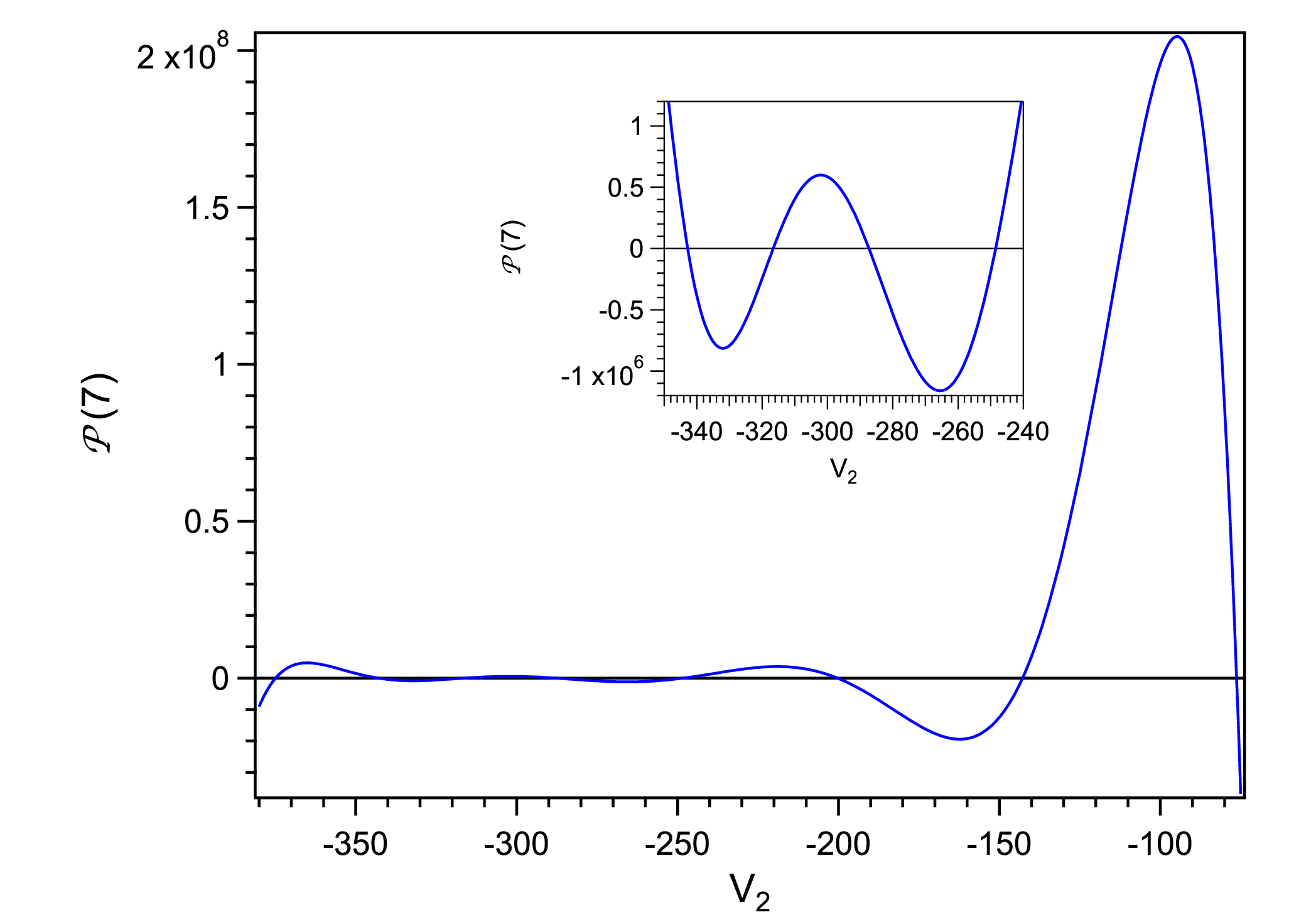}
\end{center}
\caption{Constraint polynomial for the Chen et al. generalized Manning potential 
in the {\em odd} parity case as a function of $V_2$ with fixed $V_1=0.09$, $V_3=400$, $g=0.25$, and $n=7$. 
There is the maximum number of $8$ {\em real} zeros of the constraint polynomial:
$V_2=- 374.929,\, - 342.812,\, - 316.597,\, - 287.269,\, - 248.489,\, - 200.236,\, -142.792,\, - 76.2691$. 
Polynomial eigenfunctions are shown in Fig. \ref{fgCMwfo}.
}
\label{fgcmo}
\end{figure}
\begin{figure}
\begin{center}
\includegraphics[width=14cm,clip=0,angle=0]{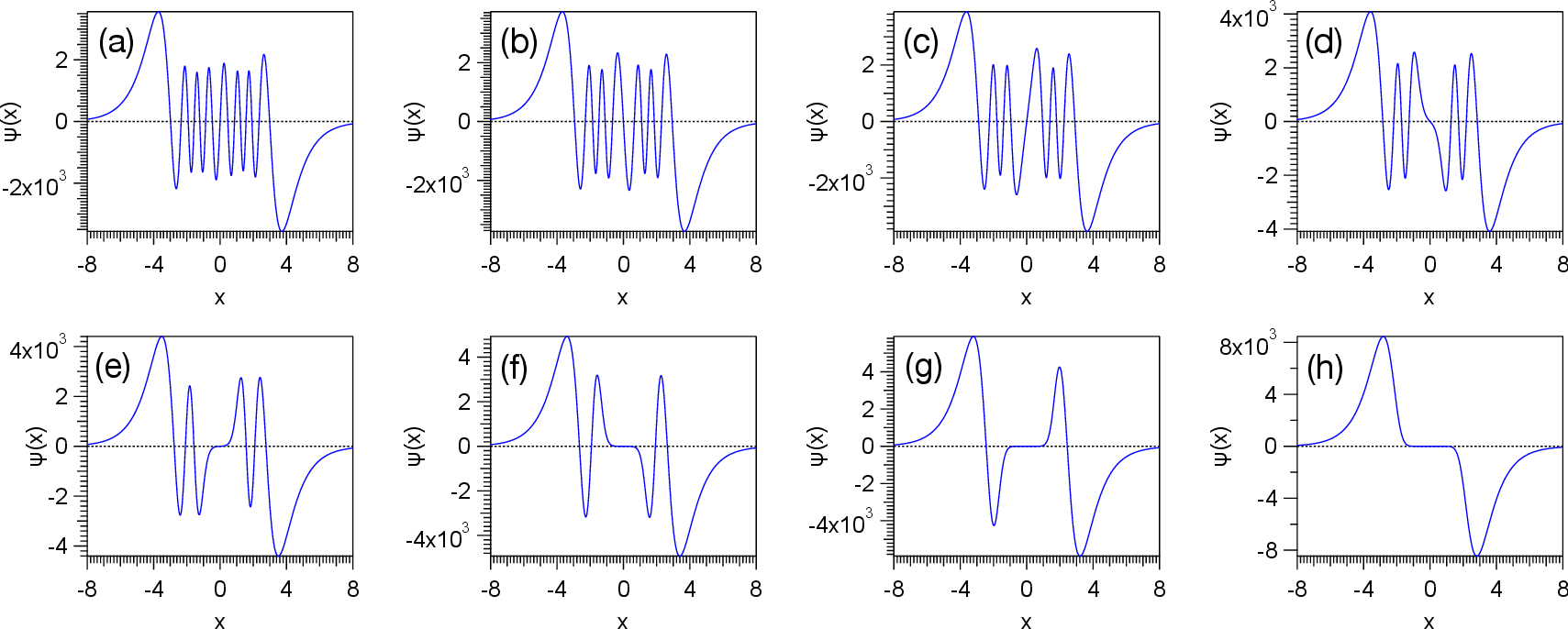}
\end{center}
\caption{Polynomial eigenfunctions for the Chen et al. generalized Manning potential 
in the {\em odd} parity case with fixed $V_1=1$, $V_3=400$, $g=0.25$, and $n=7$ for the values of 
$V_2=- 374.929,\,- 342.812,\,- 316.597,\,- 287.269,\,- 248.489,\, - 200.236,\,- 142.792,\,- 76.2691$
as in Fig. \ref{fgcmo}.
}
\label{fgCMwfo}
\end{figure}
Because $c_1$ is energy dependent, the necessary condition (\ref{bnc}),
\begin{equation}
F_1(n)=n(n-1)+ 2n(\lambda_1+\lambda_2+1)+(\lambda_1+\lambda_2)(\lambda_1+\lambda_2+1) +\frac{E+1}{4}=0,
\nonumber
\end{equation}
forces energy onto a $n$th baseline,
\begin{equation}
E_n=-1- 4[n^2+ n(2\lambda_1+2\lambda_2)+(\lambda_1+\lambda_2)(\lambda_1+\lambda_2+1) ]
 =-1- 4(n+\lambda_1+\lambda_2)(n+\lambda_1+\lambda_2+1).
\label{oCmMe}
\end{equation}
On the $n$th baseline one has in virtue of (\ref{grmchp})
\begin{eqnarray}
& F_1(k)=k(k-1)-n(n-1) +2(k-n)(\lambda_1+\lambda_2+1)>(n-k)^2>0, &
\nonumber\\
&
F_0(k)=-k(k-1) \left(2+\frac{1}{g}\right)-k\left[ 2\lambda_2+2\lambda_1+\frac72+\frac{2(\lambda_1+1)}{g}\right] 
+c_0(n), &
\nonumber\\
& F_{-1}(k)=k(k-1) \left(1+\frac{1}{g}\right) + \frac32\, k \left(1+\frac{1}{g}\right)=\frac{1+g}{2g}\, k(2k+1),& 
\label{3mFvo}
\end{eqnarray}
where
\begin{eqnarray}
c_0(n) &=& -\frac{1+g}{4g}
 \left[6\lambda_1+4\lambda_2+1+\frac{2\lambda_2 g-V_2}{1+g}-V_1-\frac{V_3}{(1+g)^2}-1
 \right.
 \nonumber\\
 && \left.\vphantom{\frac12} 
- 4(n+\lambda_1+\lambda_2)(n+\lambda_1+\lambda_2+1) \right]+\frac{\lambda_2}{g}\cdot
\nonumber
\end{eqnarray}
Fig. \ref{fgcmo} shows the constraint polynomial as a function of $V_2$
for fixed $V_1=1$, $V_3=400$, $g=0.25$, and $n=7$. Fig. \ref{fgCMwfo} displays wave functions
corresponding to the roots of the constraint polynomial of Fig. \ref{fg3}.

\subsection{Electron in Coulomb and magnetic fields and relative 
motion of two electrons 
in an external oscillator potential}
\label{sc:clmb}
After an appropriate change of parameters, (i)
the Schr\"odinger equation for electron in Coulomb and magnetic fields,
(ii) the Klein-Gordon equation for electron in Coulomb and magnetic fields,
and (iii) the three-dimensional Schr\"odinger
equation for two electrons (interacting with Coulomb potential) in an external
harmonic-oscillator potential with frequency $\omega_{ext}$ can all be shown to
have the same basic form \cite{CH}
\begin{eqnarray}
\left[\frac{1}{2}\frac{d^2}{dr^2}-\frac{g(g-1)}{2}
\frac{1}{r^2} -\frac{1}{2}\omega^2 r^2+\frac{\beta}{r}+\alpha
\right]~u(r)=0.
\label{general}
\end{eqnarray}
Here $\beta, g$ and $\omega$ ($g,\, \omega>0$) are real
parameters, and $\alpha$ is the eigenvalue of Eq. (\ref{general}) \cite{CH}.
The potential in the Schr\"odinger
equation (\ref{general}) is the only one here {\em without} a parity symmetry. 
Obviously $\lim_{r\to\infty} V(r)=\infty$.

After the change of variables: $x= \sqrt{2\omega} r$
and rescaling $\beta \to (\sqrt{2/\omega}) \beta$, Eq. (\ref{general}) becomes:
\begin{eqnarray}
\left[\frac{d^2}{dx^2}-\frac{g(g-1)}{x^2}
 -\frac{x^2}{4}+\frac{\beta}{x}+\frac{\alpha}{\omega} \right]~u(x)=0.
\label{general-x}
\end{eqnarray}
\begin{figure}
\begin{center}
\includegraphics[width=14cm,clip=0,angle=0]{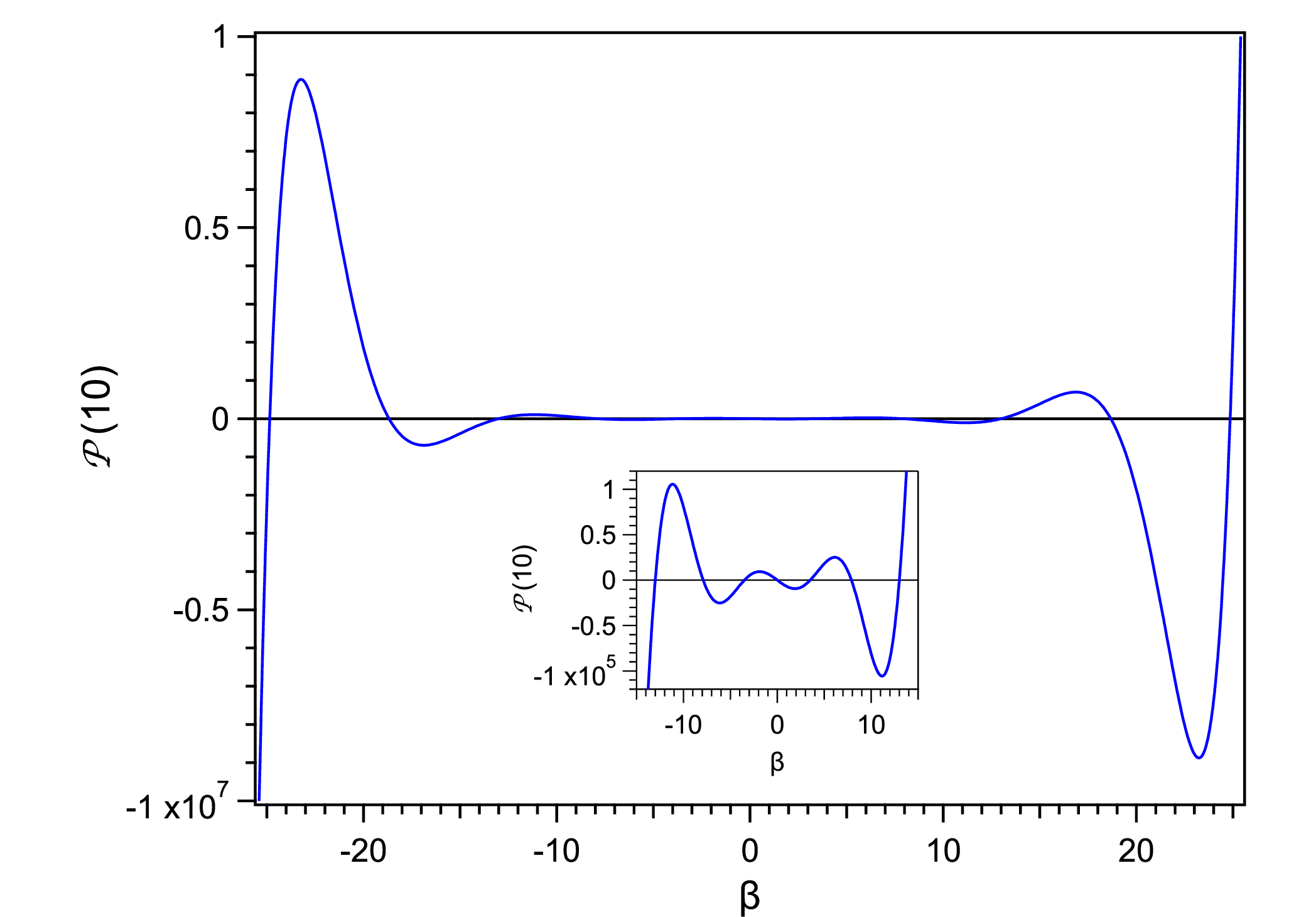}
\end{center}
\caption{Constraint polynomial as a function of $\beta$ with fixed $g=0.5$ and $n=10$ for the problem 
defined by Eq. (\ref{eqrnew}). There is the maximum number of $11$ {\em real} 
zeros of the constraint polynomial arranged symmetrically around $\beta=0$, namely
$\beta=\mp 24.8502,\, \mp 18.676,\, \mp 13.0012,\, \mp 7.89603,\, \mp 3.50671,\, 0$.
}
\label{fg4}
\end{figure}
On substituting Ansatz
\begin{eqnarray}
u(x) = x^g\exp(-x^2/4)\phi(x)
\label{eqff1}
\end{eqnarray}
into (\ref{general-x}) one obtains
\begin{eqnarray}
 \left[x\, \frac{d^2}{dx^2}+\left(2g-x^2\right)\frac{d}{dx} +
\left(\epsilon x+ \beta\right)\right] \phi(x) = 0,
\label{eqrnew}
\end{eqnarray}
where $\epsilon = \alpha/\omega -(g+1/2)$ \cite{CH}, 
which has again the form of Eq. (\ref{qese}). The Ansatz (\ref{eqff1}) yields 
a normalizable solution on the interval $x\in (0,\infty)$ 
for any polynomial $\phi(x)$, provided that $g>-1/2$.

The necessary condition $F_1(n)=-n +\epsilon=0$ forces energy onto a $n$th baseline, 
$\epsilon=n$. On the $n$th baseline one has in virtue of (\ref{grmchp})
\begin{eqnarray}
& F_1(k)=n-k, \qquad F_0(k)=\beta, \qquad F_{-1}(k)=k(k-1)+2kg.& 
\label{3CF}
\end{eqnarray}
The choice of $\beta$ as the spectral parameter is in virtue of $F_0(k)=\beta$ unavoidable here.
Being a linear function, $F_1(k)$ has for each $n$ only single zero. Hence the conditions (\ref{th1nsc})
are satisfied and there can always be only a unique polynomial solution.

\begin{figure}
\begin{center}
\includegraphics[width=14cm,clip=0,angle=0]{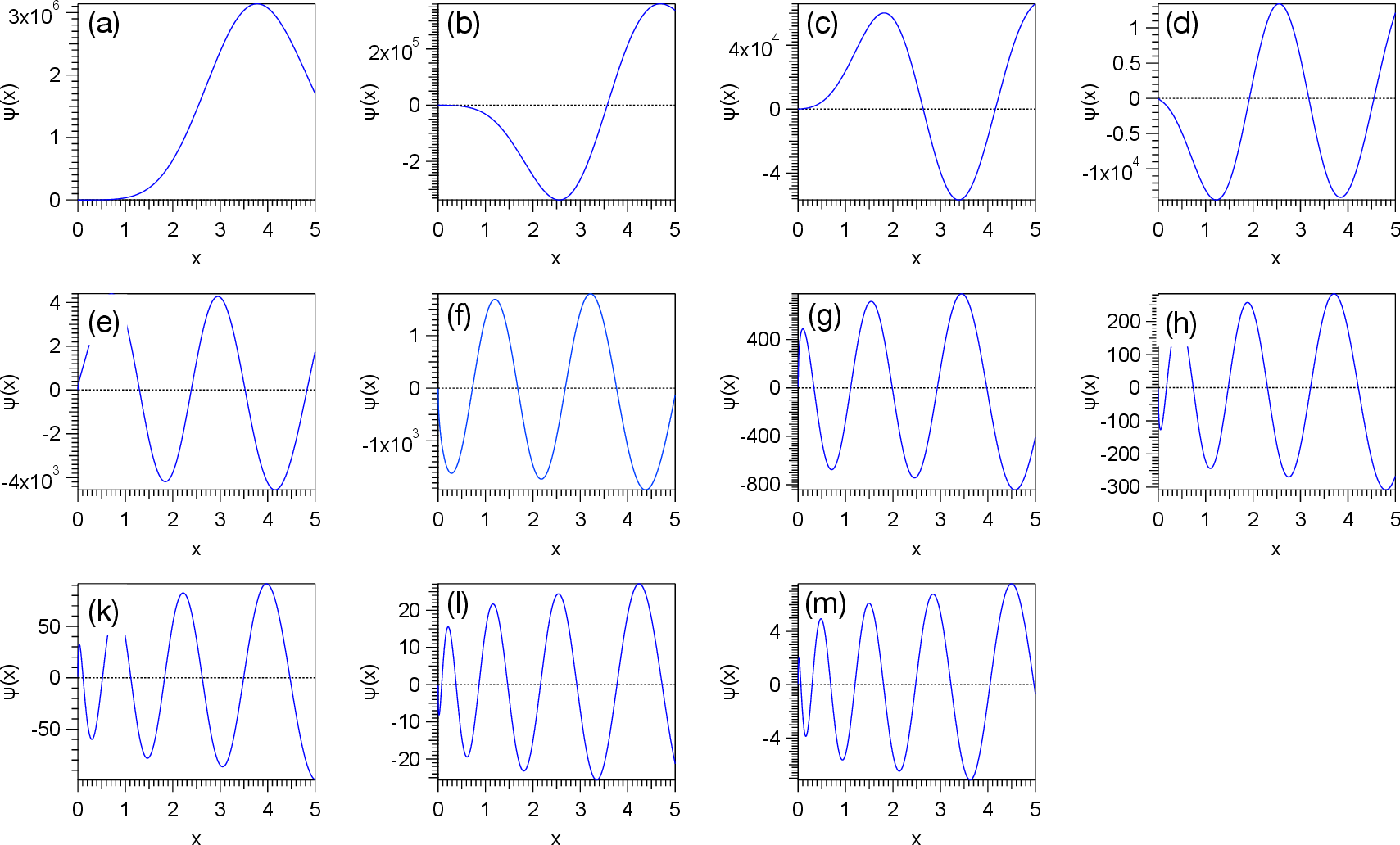}
\end{center}
\caption{Wave functions given by the Ansatz (\ref{eqff1}) for the roots of the constraint 
polynomial shown in Fig. \ref{fg4} ordered from the lowest till the highest one.
}
\label{fgclmb}
\end{figure}

The resulting equation is {\em symmetric}
under simultaneous transformation $\beta\to -\beta$ and $x\to -x$. The latter implies that if 
$\phi(x)$ solves (\ref{eqrnew}) for some $\beta_0$, then also $\phi(-x)$ is a solution 
of Eq. (\ref{eqrnew}), but with the eigenvalue $-\beta_0$. In particular, 
the eigenvalue $\beta=0$ is possible only for $n$ even if all the roots of ${\cal P}(n)$
are simple [${\cal P}(n)$ has $n+1$ roots].
The latter is explicitly manifested in the distribution of eigenvalues in Fig. \ref{fg4}. 
Fig. \ref{fgclmb} displays wave functions
corresponding to the roots of the constraint polynomial of Fig. \ref{fg4}.

\section{Examples of only $F_0(n)$ depending on energy}
\label{sc:wop}

\subsection{The hyperbolic Razavy potential}
\label{sc:hRp}
In this section we examine parity invariant potential (cf. Eq. (2.6) of Ref. \cite{Rzv})
\begin{equation}
V(x) =\frac18\, \xi^2 [\cosh(4x)-1]- (M+1)\xi \cosh(2x)=\frac14\, \xi^2 \sinh^2 (2x) - (M+1)\xi \cosh(2x),
\label{hRp} 
\end{equation}
$\lim_{|x|\to\infty} V(x)=\infty$. The Ansatz \cite{HaS}
\begin{equation}
\psi(x)=\exp\left(-\frac{\xi}{4}\cosh 2x\right) \left(\cosh^\alpha x\right) 
 \left(\sinh^\beta x\right)\, \phi(x)
\label{hRA}
\end{equation}
transforms the Schr\"odinger equation in virtue of (\ref{QetahR}) into
\begin{eqnarray}
&\left[ \vphantom{\frac12} d_x^2 + \left(-\xi \sinh 2x+2\alpha \tanh x+2\beta \coth x \right)d_x
+E+(\alpha+\beta)^2 \right.&
\nonumber\\
&
\left.
+ M \xi \cosh(2x) -2 \xi(\alpha\sinh^2 x+\beta \cosh^2x)\right]\phi=0,&
\label{etappr}
\end{eqnarray}
where $\alpha(\alpha-1)=\beta(\beta-1)=0$ (i.e. $\alpha\in\{0,1\}$, $\beta\in\{0,1\}$).
Assuming the substitution $z=\cosh^2 x$, the Ansatz (\ref{hRA}) yields a normalizable solution 
on the interval $x\in (-\infty,\infty)$ for any polynomial $\phi(z)$.
\begin{figure}
\begin{center}
\includegraphics[width=14cm,clip=0,angle=0]{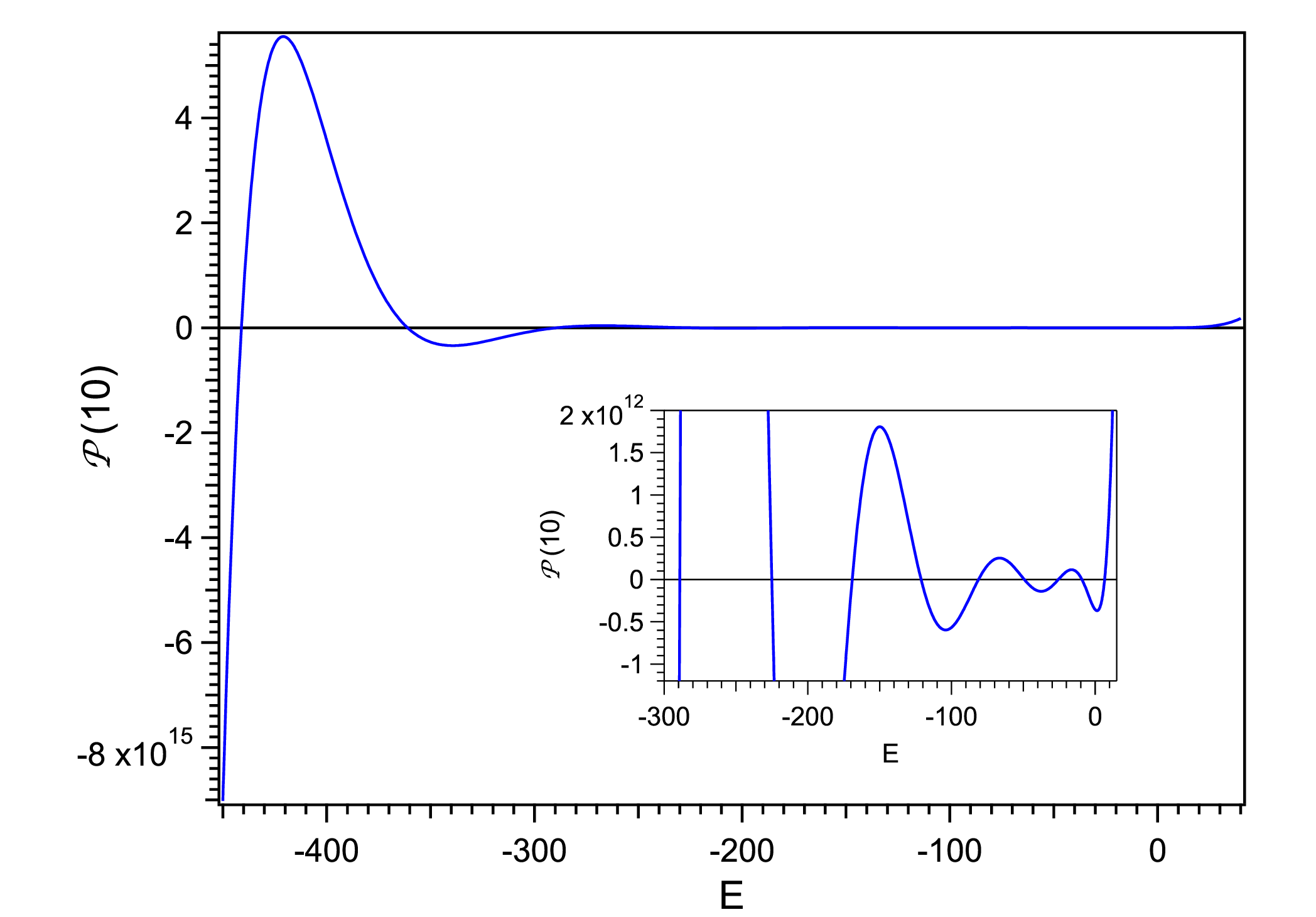}
\end{center}
\caption{Constraint polynomial for the hyperbolic Razavy potential as a function of energy $E$
with fixed $\xi=0.5$, $\alpha=0$, odd parity $\beta=1$, and $n=10$. There is the 
maximum number of $11$ {\em simple real} roots $E=-441.066,\,-361.073,\, -289.084,\,-225.099,\, -169.121,
\, -121.157,\,-81.2206,\, -49.3476,\, -25.6452,\newline \, -9.23983,\, 6.55323$.
}
\label{fg5}
\end{figure}
\begin{figure}
\begin{center}
\includegraphics[width=14cm,clip=0,angle=0]{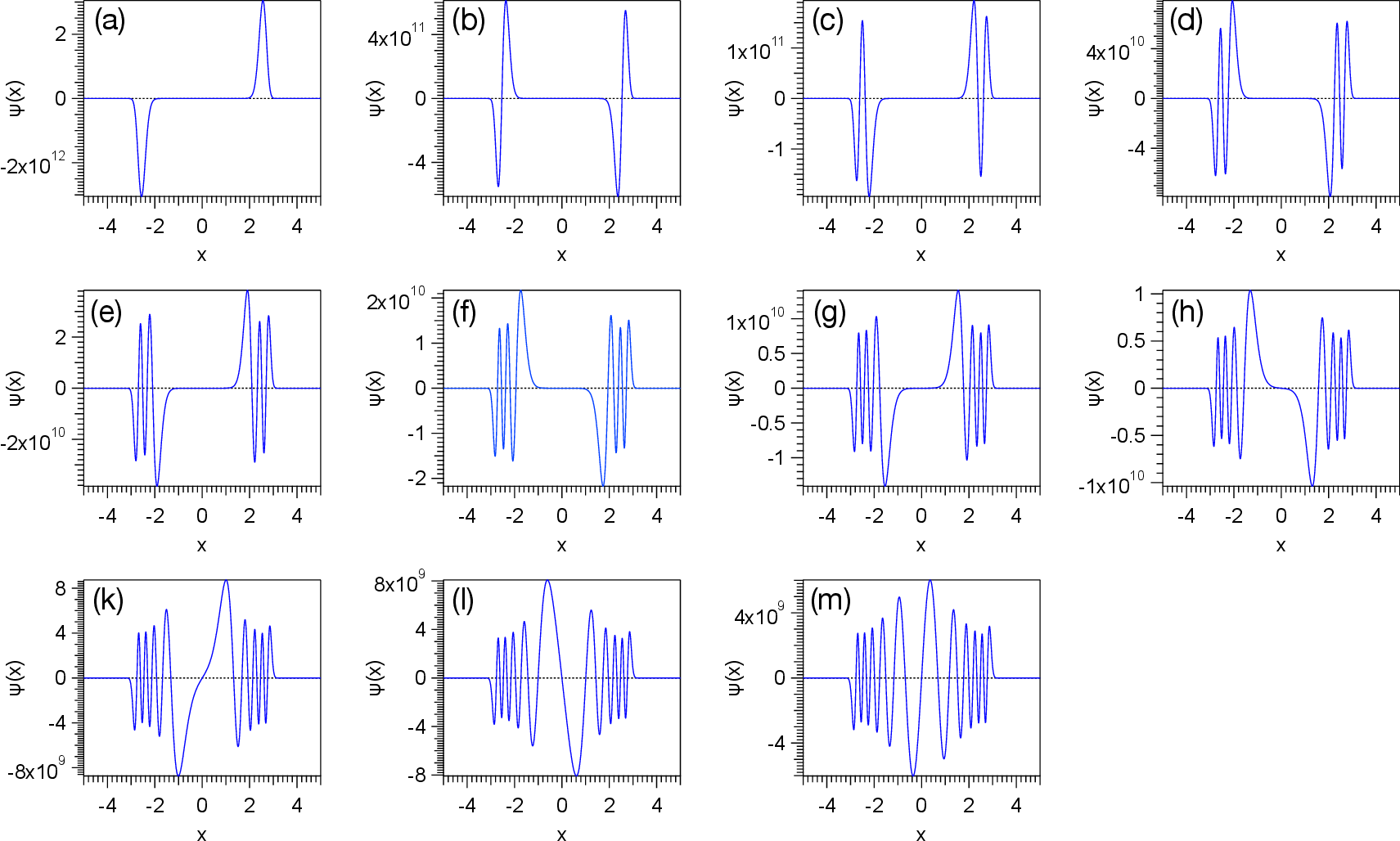}
\end{center}
\caption{Odd parity polynomial eigenfunctions for the hyperbolic Razavy potential 
given by the Ansatz (\ref{hRA}) with $\phi$ there being a polynomial in $z=\cosh^2x$ 
for fixed $\xi=0.5$ and $n=10$ for the $11$ {\em simple real} roots 
$E=-441.066,\,-361.073,\, -289.084,\,-225.099,\, -169.121,
\, -121.157,\,-81.2206,\, -49.3476,\, -25.6452,\newline \, -9.23983,\, 6.55323$ 
of the constraint polynomial of Fig. \ref{fg5}.
}
\label{fgRwf}
\end{figure}
The substitution $z=\cosh^2 x$ transforms 
the differential operator in (\ref{etappr}) in virtue of (\ref{chss}) into
\begin{eqnarray}
& 4z(z-1)\, d_z^2+\left[-4\xi z^2 +4(\alpha+\beta+\xi+1)z -2(2\alpha+1)\right]\, d_z &
\nonumber\\
& +\left[2\xi (M-\alpha-\beta)z+E+ (\alpha+\beta)^2 -\xi(M-2\alpha)\right], & 
\nonumber
\end{eqnarray}
which is (\ref{qese}) with
\begin{eqnarray}
& a_2 = 4,\qquad a_1 =-4,& 
\nonumber\\
& b_2 = -4\xi,\qquad b_1 = 4(\alpha+\beta+\xi+1),\qquad b_0 = -2(2\alpha+1),& 
\nonumber\\
& c_1 = 2\xi(M-\alpha-\beta),\qquad c_0 = E+(\alpha+\beta)^2+\xi(2\alpha-M).& 
\label{Rzvcf} 
\end{eqnarray}
The necessary condition $F_1(n)=-4n\xi+2\xi(M-\alpha-\beta)=0$ is solved by
\begin{equation}
M = 2n+\alpha+\beta.
\nonumber
\end{equation}
On the $n$th baseline one has in virtue of (\ref{grmchp})
\begin{eqnarray}
& F_1(k)=4\xi (n-k), \qquad F_0(k)=4k (k+\alpha+\beta +\xi)+c_0(n), &
\nonumber\\
& F_{-1}(k)=-2k(2k-1+2\alpha),& 
\label{RzvF}
\end{eqnarray}
where
\begin{equation}
c_0(n)= E+(\alpha+\beta)^2- \xi(2n+\beta-\alpha).
\label{c0rzv} 
\end{equation}
The even (odd) parity solutions given by the Ansatz (\ref{hRA}) correspond to $\beta=0$ ($\beta=1$).

It turned out straightforward to reproduce energy levels $E_{n,\alpha,\beta}$ for $n,\alpha,\beta=0,1$ of 
the hyperbolic Razavy potential given in Eqs. (45), (47), (49), (52), (56), (58), (60), (64), (65) of \cite{HaS}. 
Note in passing that when comparing our energy levels $E_{n,\alpha,\beta}$ against those in Ref. \cite{HaS}
one has to interchange $\alpha$ and $\beta$.
Fig. \ref{fg5} shows constraint polynomial as a function of $E$ 
for fixed $\xi=0.5$, $\alpha=0$, odd parity $\beta=1$, and $n=10$.
Fig. \ref{fgRwf} displays wave functions
corresponding to the roots of the constraint polynomial of Fig. \ref{fg5}.

\subsection{A double sinh-Gordon system}
\label{sc:dshg}
The double sinh-Gordon (DSHG) parity invariant system (also called the 
bistable Razavy potential \cite{BP}) is characterized by the potential
\begin{equation}
V (x) = [\xi\cosh(2x) -M]^2,
\label{dshgp}
\end{equation}
where $\xi$ and $M$ are positive real parameters and 
$\lim_{|x|\to\infty} V(x)=\infty$. The potential is one of the few double well problems 
in quantum mechanics which is QES. 
\begin{figure}
\begin{center}
\includegraphics[width=14cm,clip=0,angle=0]{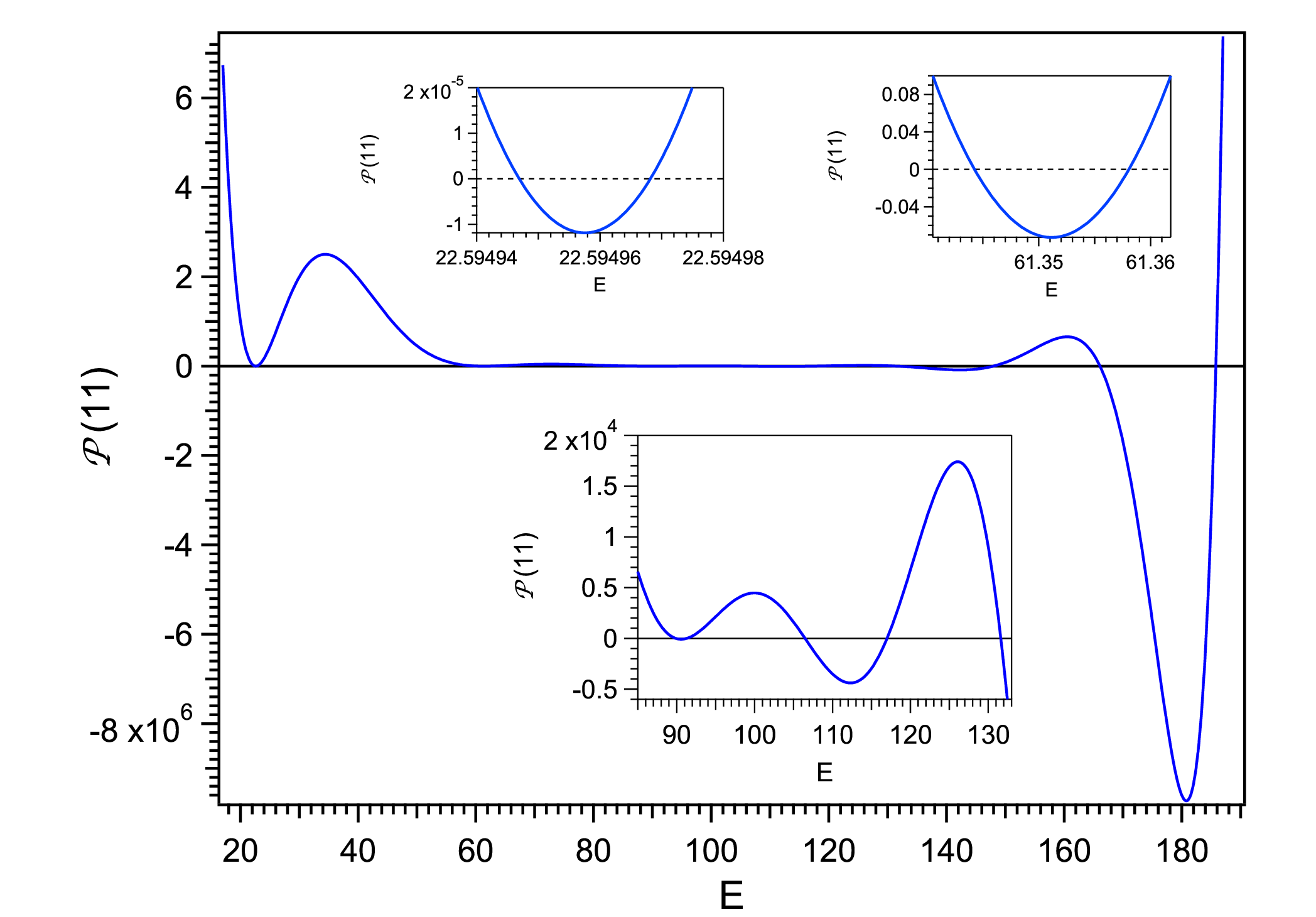}
\end{center}
\caption{Constraint polynomial for the DSHG with $\xi=2$ 
on the $11$th baseline corresponding to $M=12$ is shown to have the maximum number of $12$ {\em simple real} roots 
$E= 22.59494691,\,22.59496818,\,61.34425227,\,61.35805469,\,89.87448537,\,91.28081517,\,106.4782162,\,
\\
117.0076415,\,131.6165721,\,147.9807662,\,166.0915272,\,185.7777543$ 
reproducing the results of Tab. 3 of Ref. \cite{BP}.
}
\label{fg6}
\end{figure}
\begin{figure}
\begin{center}
\includegraphics[width=14cm,clip=0,angle=0]{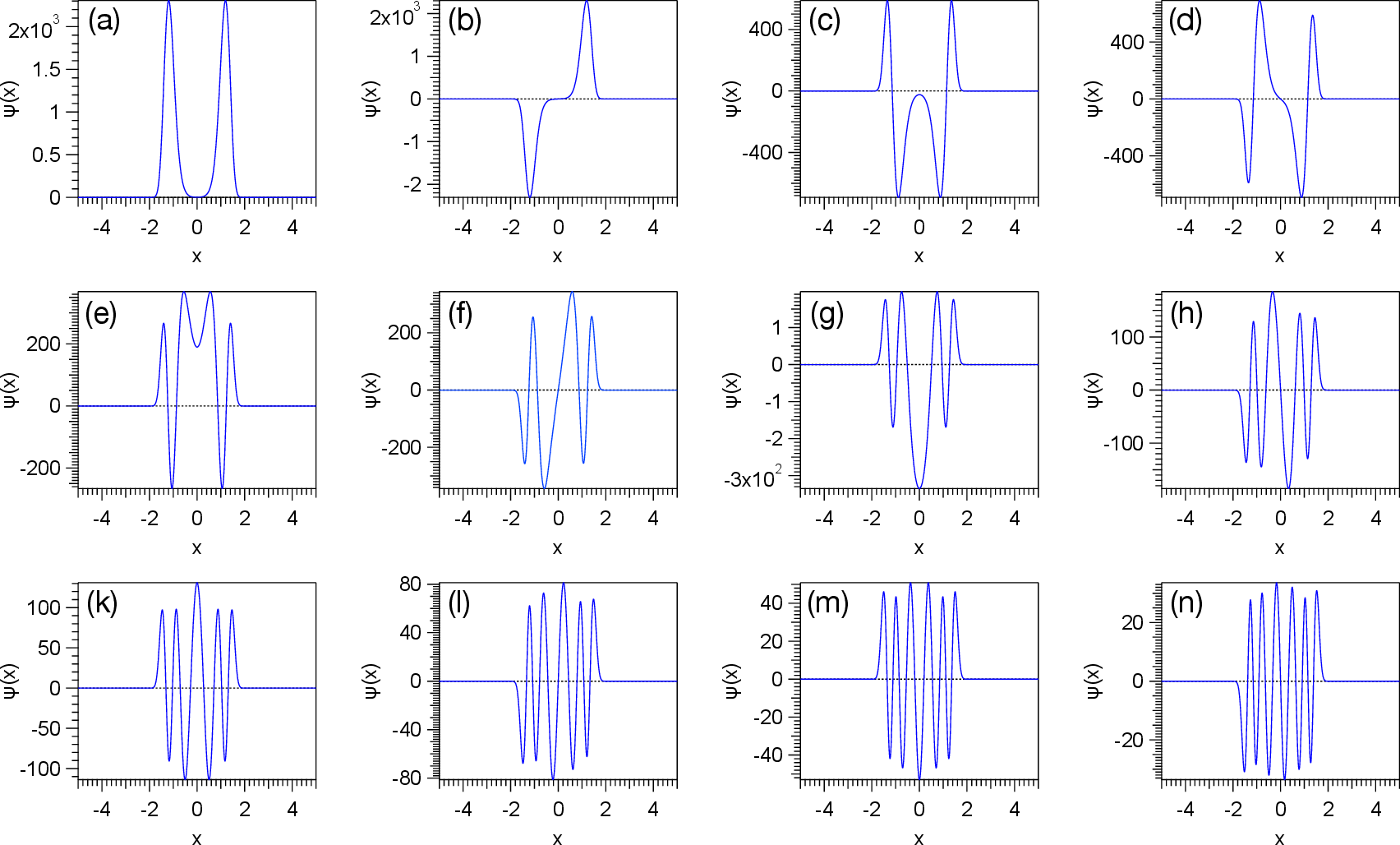}
\end{center}
\caption{Interlaced even and odd parity polynomial eigenfunctions for the DSHG 
given by the Ansatz (\ref{dshgA}) with fixed $\xi=2$ and $n=11$ corresponding to the twelve {\em simple real} roots 
$E= 22.59494691,\,22.59496818,\,61.34425227,\,61.35805469,\,89.87448537,\,91.28081517,\,106.4782162,\,
\\
117.0076415,\,131.6165721,\,147.9807662,\,166.0915272,\,185.7777543$ 
of the constraint polynomial of Fig. \ref{fg6}.
}
\label{fgdsg}
\end{figure}

The change of independent variable $z= e^{2x}$ and
\begin{equation}
\psi(z)=z^{\tfrac{1-M}{2}}\, \exp\left[ -\frac{\xi}{4} \left(z+\frac{1}{z} \right) \right] \phi(z)
\label{dshgA}
\end{equation}
transform the Schr\"odinger
equation (\ref{SE}) into (\ref{qese}) with \cite{BP} (cf. Appendix \ref{sc:gtr})
\begin{eqnarray}
& a_2 = 4,\qquad a_1 = 0,& 
\nonumber\\
& b_2 =-2\xi,\qquad b_1 = 8-4M,\qquad b_0 = 2\xi,& 
\nonumber\\
& c_1 = 2\xi(M-1),\qquad c_0 =E+1-2M-\xi^2.& 
\nonumber
\end{eqnarray}
The Ansatz (\ref{dshgA}) yields normalizable solutions 
on the interval $x\in (-\infty,\infty)$ for any polynomial $\phi(z)$.

The baseline condition $F_1(n)=-2n\xi+2\xi(M-1)=0$ is satisfied by $n=M-1$. 
Hence the Ansatz (\ref{dshgA}) will comprise polynomial powers of $z$ between 
$z^{-n/2}=e^{-nx}$ up to $z^{n/2}=e^{nx}$. On the $n$th baseline one has in virtue of (\ref{grmchp})
\begin{eqnarray}
& F_1(k)=2\xi(n-k), \qquad F_0(k)=-4k(n-k) +c_0(n), \qquad F_{-1}(k)=2k\xi,&
\label{DSFpm}
\end{eqnarray}
where $c_0(n)= E-\xi^2-2n-1$.

It turned out straightforward to reproduce energy levels for the double sinh-Gordon system in 
Tab. 2, 3 of \cite{BP}, which contain numerous energy levels and the energy levels 
splitting with $\xi=2$ and $M$ between $1$ and $12$.
Fig. \ref{fg6} shows constraint polynomial for a double sinh-Gordon system for $n=11$, 
corresponding to $\xi=2$ and $M=12$ of Ref. \cite{BP}.
Fig. \ref{fgdsg} displays wave functions
corresponding to the roots of the constraint polynomial of Fig. \ref{fg6}.

Because $V(x)$ in (\ref{dshgp}) has even parity, the solutions has to have definite parity. 
Yet it is difficult to identify the parity of solutions on using the Ansatz (\ref{dshgA}).
The latter will be answered in Sec. \ref{sc:pdshg} on using the Ansatz Eq. (\ref{hRAd}) 
for the special case when $\alpha(\alpha-1)=\beta(\beta-1)\equiv 0$
[cf. the condition (\ref{qdrp})], i.e. when $\alpha\in\{0,1\}$, $\beta\in\{0,1\}$.

\begin{figure}
\begin{center}
\includegraphics[width=14cm,clip=0,angle=0]{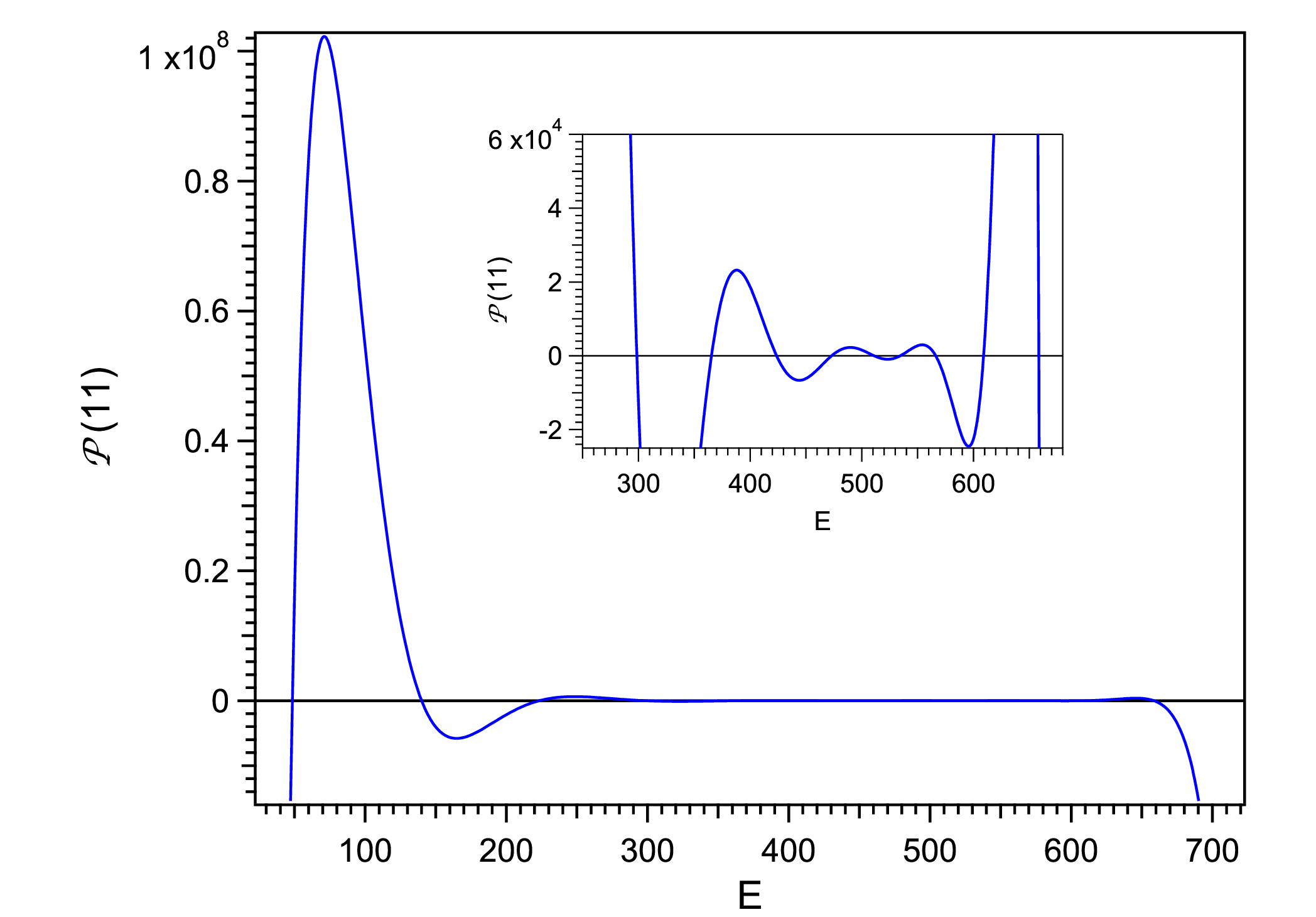}
\end{center}
\caption{
Constraint polynomial for the perturbed DSHG on the $n=11$th baseline 
for $\alpha=2$, $\beta=0$ (i.e. even parity states), 
and $\xi=2$, corresponding to $g(g+1)=2$ and $h(h+1)\equiv 0$ in the respective numerators of the potential (\ref{dshgpp}). 
There is the maximal number of twelve simple real zeros 
$E= 48.5067,\, 140.039,\, 223.425,\, 298.596,\, 365.435,\, 423.725,\, 472.987,\, 
 511.035,\, 534.418,\, 566.233,\,\newline 609.075,\, 658.526$.
 }
\label{fgpdshge}
\end{figure}
\begin{figure}
\begin{center}
\includegraphics[width=14cm,clip=0,angle=0]{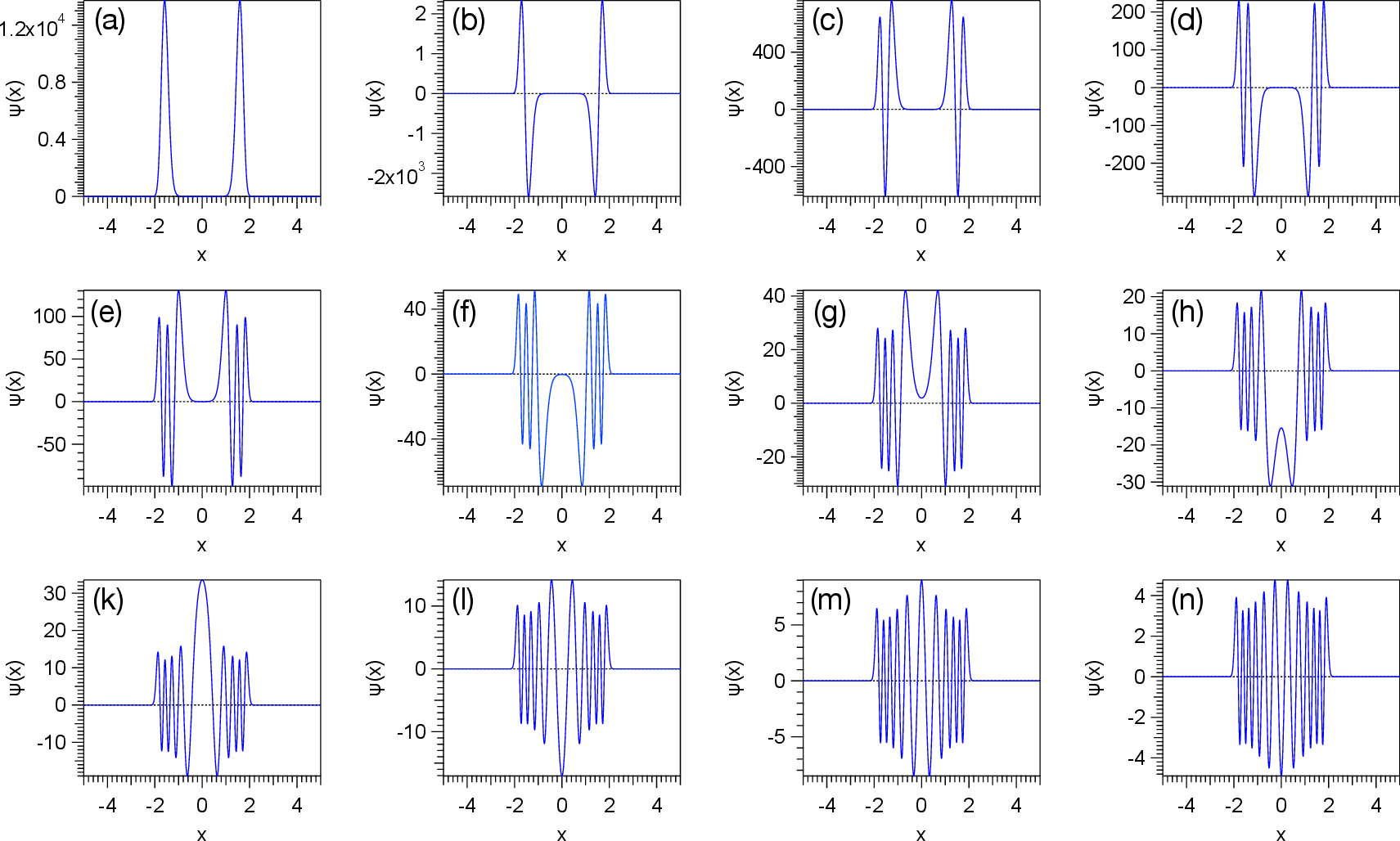}
\end{center}
\caption{
Even parity polynomial eigenfunctions for the perturbed DSHG 
given by the Ansatz (\ref{hRAd}) with $\phi$ there being a polynomial in $z=\cosh^2x$ and
fixed $\alpha=2$, $\beta=0$, $\xi=2$, corresponding to the twelve {\em simple real} roots 
of the constraint polynomial of Fig. \ref{fgpdshge}.
}
\label{fgwfdshge}
\end{figure}

\subsection{A perturbed double sinh-Gordon system}
\label{sc:pdshg}
Khare and Mandal \cite{KM} showed that after adding a parity invariant perturbation 
\begin{equation}
V_p=-\frac{g(g+1)}{\cosh^2 x}+\frac{h(h+1)}{\sinh^2 x}
\label{dshgpp}
\end{equation}
term to the DSHG potential (\ref{dshgp}), 
the resulting potential is still QES potential (cf. Eq. (41) of Ref. \cite{KM}). 
Because $\sinh^2 x$ is singular at the origin, the singularity is usually tamed by imposing the 
restriction $-1<h\le 0$ on $h\in\mathbb{R}$ \cite{KM}, which limits the product $h(h+1)\in (-0.25,0)$. 
(For $h(h+1)\le -0.25$ one has the familiar textbook ``{\em fall to the center}'' - a particle
falls in the origin and one cannot prevent the spectrum from collapse by any means \cite{LL3,Zn0}.)
On the other hand, $\cosh^2 x$ is regular at the origin and the potential 
parameter $g\in\mathbb{R}$ is unrestricted.
\begin{figure}
\begin{center}
\includegraphics[width=14cm,clip=0,angle=0]{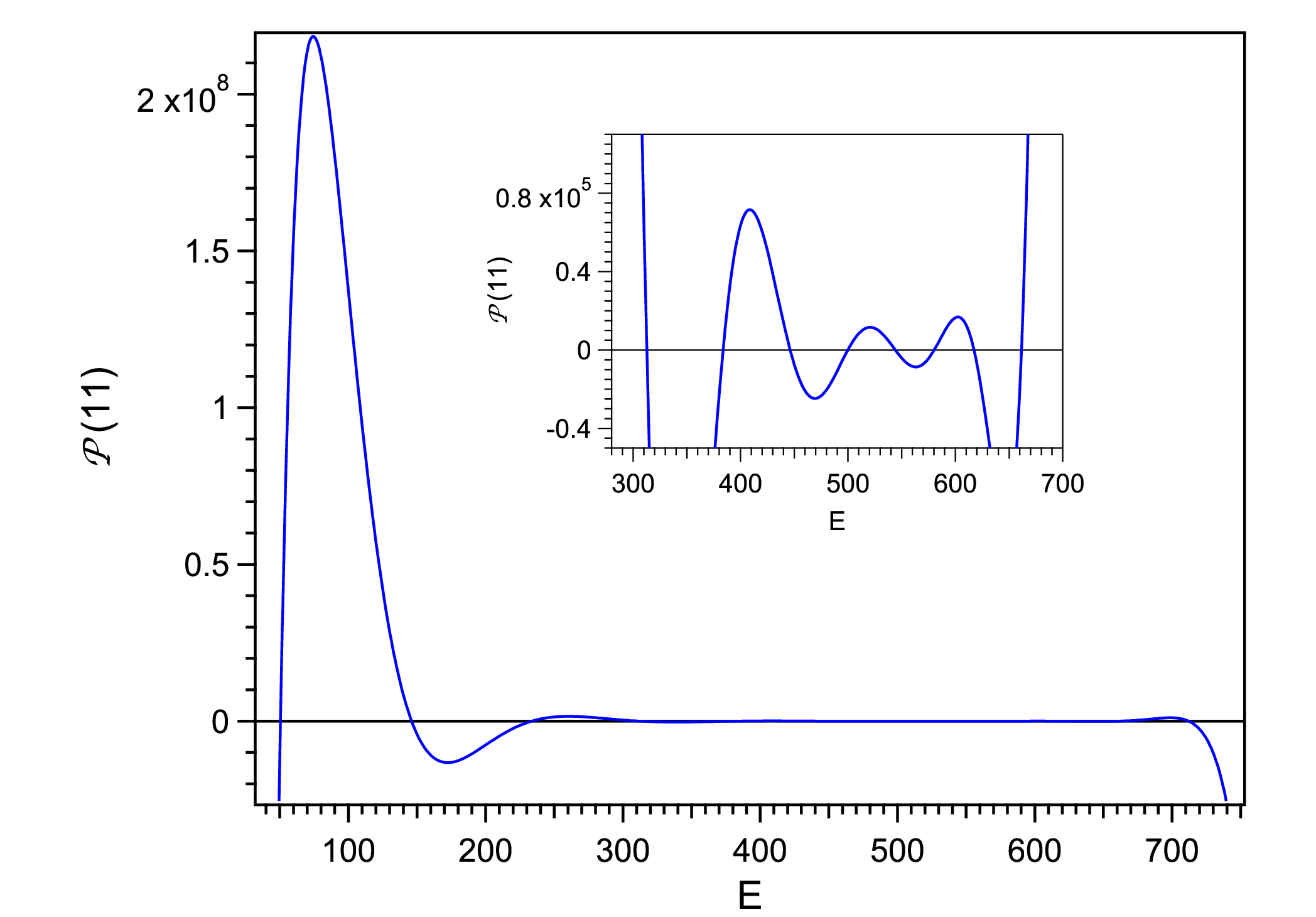}
\end{center}
\caption{
Constraint polynomial for the perturbed DSHG on the $n=11$th baseline for $\alpha=2$, $\beta=1$ (i.e. odd parity states), 
and $\xi=2$, corresponding to $g(g+1)=2$ and $h(h+1)\equiv 0$ in the respective numerators of 
the potential (\ref{dshgpp}). There is the maximal number of twelve simple real zeros 
$E= 50.5262,\, 146.083,\, 233.507,\, 312.742,\, 383.69,\, 446.18,\, 499.874,\, 544.126,\, 
580.222,\, 617.352,\,\newline 661.546,\, 712.152$
}
\label{fgpdshgo}
\end{figure}
\begin{figure}
\begin{center}
\includegraphics[width=14cm,clip=0,angle=0]{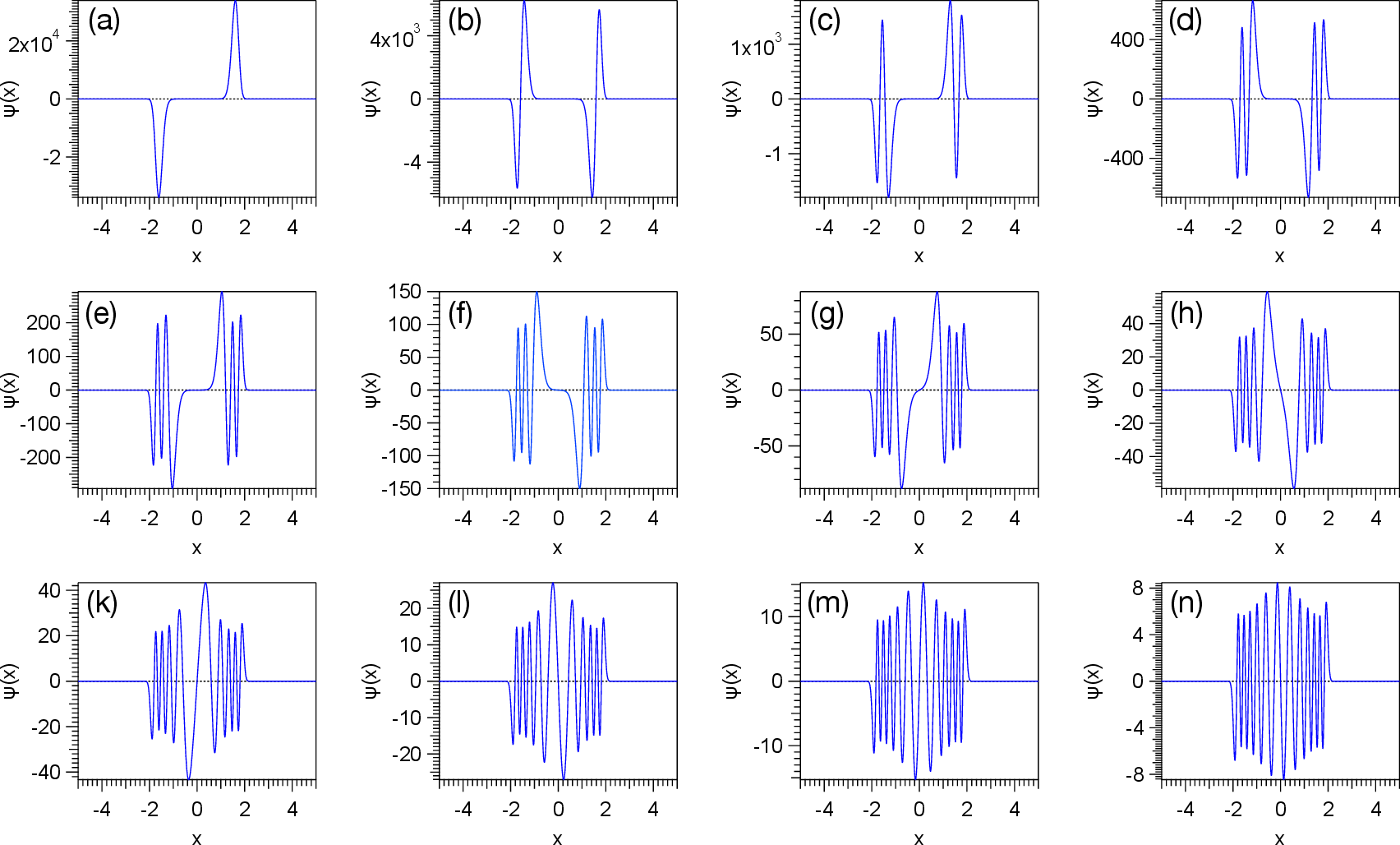}
\end{center}
\caption{
Odd parity polynomial eigenfunctions for the perturbed DSHG 
given by the Ansatz (\ref{hRAd}) with $\phi$ there being a polynomial in $z=\cosh^2x$
and fixed $\alpha=2$, $\beta=1$, $\xi=2$,
corresponding to the twelve {\em simple real} roots 
of the constraint polynomial of Fig. \ref{fgpdshgo}.
}
\label{fgwfdshgo}
\end{figure}

The Ansatz 
\begin{equation}
\psi(x)=\exp\left(-\frac{\xi}{2}\cosh 2x\right) \left(\cosh^\alpha x\right) 
 \left(\sinh^\beta x\right)\, \phi(x),
\label{hRAd}
\end{equation}
which differs from that of Eq. (\ref{hRA}) in $\xi\to 2\xi$, 
transforms the Schr\"odinger equation (\ref{SE}) in virtue of (\ref{Qeta}) into
\begin{eqnarray}
&\left[ \vphantom{\frac12} d_x^2 + 2\left(-\xi \sinh 2x+\alpha \tanh x+\beta \coth x \right)d_x
+E-M^2-\xi^2+(\alpha+\beta)^2 
\right.&
\nonumber\\
&
\left.
 +\, 2 \xi(2\alpha-M+1) +4 \xi(M-\alpha-\beta-1)\cosh^2x \right]\phi=0,&
\label{etapp}
\end{eqnarray}
provided that 
\begin{equation}
\alpha(\alpha-1)=g(g+1),\qquad \beta(\beta-1)=h(h+1).
\label{qdrp}
\end{equation}
The condition determines for a given $g$ and $h$ a {\em quadruplet} of 
energy values characterized by $\alpha=g+1,\, -g$ and $\beta=h+1,\, -h$.
The solutions expressed by the Ansatz (\ref{hRAd}) are normalizable
on the interval $x\in(-\infty,\infty)$ for any polynomial $\phi(x)$.

Similarly to the hyperbolic Razavy potential of Sec. \ref{sc:hRp}, either substitution
$z=\cosh^2x$ or $z=\sinh^2 x$ transforms the Schr\"odinger equation into (\ref{qese}).
With $z=\cosh^2x$, Eq. (\ref{etapp}) is transformed in virtue of (\ref{chss})
into (\ref{qese}) with \cite{KM}
\begin{eqnarray}
& a_2 = 4,\qquad a_1 = -4,& 
\nonumber\\
& b_2 = -8\xi,\qquad b_1 = 4(\alpha+\beta+2\xi+1),\qquad b_0 = -2(2\alpha+1),& 
\nonumber\\
& c_1= 4\xi(M-\alpha-\beta-1),\qquad 
c_0 = E -M^2- \xi^2+(\alpha+\beta)^2+2 \xi(2\alpha-M+1).& 
\nonumber
\end{eqnarray}
Note for consistency that the $a_j$ and $b_j$ coefficients here differ from those in Eq. (\ref{Rzvcf})
by the substitution $\xi \to 2\xi$.
 
The necessary condition $F_1(n)=-8n\xi+ 4\xi(M-\alpha-\beta-1)=0$ is solved by
\begin{equation}
M =2n+\alpha+\beta+1. 
\label{dshgb}
\end{equation}
On the $n$th baseline one has in virtue of (\ref{grmchp})
\begin{eqnarray}
& F_1(k)=8\xi(n-k), \qquad F_0(k)= 4k(k+\alpha+\beta+2\xi)+c_0(n), &
\nonumber\\
& F_{-1}(k)=-2k(2k-1+2\alpha),& 
\label{KMFp}
\end{eqnarray}
where
\begin{equation}
c_0(n) = E -(2n+1)(2n+1+ 2\alpha+2\beta)- \xi^2 +2\xi(\alpha-\beta-2n).
\label{c0dsh} 
\end{equation}
Being a linear function, $F_1(k)$ in Eq. (\ref{KMFp}) has for each $n$ only single zero. 
Hence the conditions (\ref{th1nsc}) are satisfied and there can always be only a unique polynomial solution.

The parity of solutions is controlled by the value of $\beta$: for even (odd) parity solutions
$\beta$ has to be an even (odd) integer. Yet $\beta$ {\em need not} be an integer here [cf. Eq. (\ref{qdrp})],
in which case one has solutions in a parity invariant system without any definite parity.
This weird and paradoxical behaviour has its origin in the well-known fact that 
for $h(h+1)\in (-0.25,0)$ the potential problem involving the perturbation $V_p$ can only be well-defined
(i) on the semi-infinite interval $x\in(0,\infty)$ and (ii) after imposing
boundary condition $\lim_{x\downarrow 0} \psi(x)/\sqrt{x}=0$ at $x=0$
\cite{LL3,Zn0}. In what follows we do not want to go into the technical details here and 
plot wave functions merely for the case $h(h+1)\equiv 0$.
Fig. \ref{fgpdshge} shows constraint polynomial for the perturbed DSHG on the $n=11$th baseline
with fixed $\alpha=2$, $\beta=0$, and $\xi=2$, corresponding to $g(g+1)=2$ and $h(h+1)=0$ in the respective numerators 
of the potential (\ref{dshgpp}). Fig. \ref{fgwfdshge} displays even parity polynomial eigenfunctions 
of the perturbed DSHG corresponding to the twelve {\em simple real} roots 
of the constraint polynomial of Fig. \ref{fgpdshge}.
Similarly, Fig. \ref{fgpdshgo} shows constraint polynomial for the perturbed DSHG 
on the $n=11$th baseline with fixed $\alpha=2$, $\beta=1$, and $\xi=2$, 
again corresponding to $g(g+1)=2$ and $h(h+1)=0$ in the respective numerators 
of the potential (\ref{dshgpp}). Fig. \ref{fgwfdshgo} displays the odd parity polynomial eigenfunctions 
of the perturbed DSHG corresponding to the twelve {\em simple real} roots 
of the constraint polynomial of Fig. \ref{fgpdshgo}.

At the end of this section we want to show that the Ansatz (\ref{hRAd}) can be used to disentangle parity 
of the algebraic spectrum of the unperturbed DSHG parity invariant system of Sec. \ref{sc:dshg}. The unperturbed DSHG
is covered by the Ansatz (\ref{hRAd}) as a special case for $\alpha(\alpha-1)=\beta(\beta-1)\equiv 0$
[cf. the condition (\ref{qdrp})], i.e. when $\alpha\in\{0,1\}$, $\beta\in\{0,1\}$. 
With $z=\cosh^2x$, the baseline condition (\ref{dshgb}) can be satisfied for $M=12$ provided that $n=5$
and either (i) $\alpha=1$ and $\beta=0$ yielding {\em even} parity solutions, 
or (ii) $\alpha=0$ and $\beta=1$ yielding {\em odd} parity solutions. One finds, without any need of plotting 
wave functions as in Fig. \ref{fgdsg}, that the 
eigenvalues on the $n=11$ baseline in the caption of Fig. \ref{fg6} correspond to interlaced even and odd 
parity solution, beginning with the lowest energy {\em even} parity state.

\section{Discussion}
\label{sec:disc}
Earlier approaches in determining exact solutions of the QES solvable models discussed here 
employed without exception the functional Bethe Ansatz method \cite{HaS,BP,BPdw}. However, the latter requires
a whole set of of $n$ coupled algebraic equations to be solved simultaneously.
For instance, the use of Bethe Ansatz allows to write eigenvalues for the hyperbolic 
Razavy potential formally as
\begin{equation}
E_{n,\alpha,\beta} = 4\xi \sum_{i=1}^n z_i-(\alpha+\beta)^2+\xi(\alpha-\beta) 
- 4n \left(n+\alpha+\beta+\frac{\xi}{2}\right),
\nonumber
\end{equation}
yet the roots $z_i$ remain to be determined by a set of $n$ coupled equations of the Bethe Ansatz.
(Note in passing that the range of applicability of the functional Bethe Ansatz method \cite{Zh} has been 
recently expanded - cf. Theorem 4 and Remark 9 of Ref. \cite{AMcp}.)
For general values of $n$ solving the system of Bethe Ansatz equation is
difficult, and one must resort to numerical methods of solving a coupled set of equations \cite{FAS}. 
Contrary to that, the gradation slicing was shown to be universal and easily applicable algorithmic
recursive approach for obtaining polynomial solutions.

The list of potential considered here is far from being exhaustive.
For a complete list of the potentials 
that can be brought to the form (\ref{qese}) see recent work by
Turbiner \cite{T16,TPR16} and Ishkhanyan \cite{Ishk18,IshcH,Ishk17}.
For example, both Xie and Chen et al. modified Manning potentials with three parameters are nothing but
particular representative of $(1/2,1/2,0)$ class considered in Ref. \cite{Ishk18}.
The list includes QES potentials associated with the P\"oschl-Teller potential, the generalized 
P\"oschl-Teller potential, the Scarf potential, sextic oscillator and an anharmonic 
oscillator potential \cite{KK}, and many further
potentials, such as a number of spherically symmetric potentials \cite{PBss}
including a non-polynomial oscillator defined as 
\begin{equation}
V(r)= r^2+\frac{\alpha r^2}{1+\beta r^2},
\nonumber
\end{equation}
the screened Coulomb potential defined by, 
\begin{equation}
V(r)=\frac{\lambda}{r}+\frac{\delta}{r+\kappa},\qquad \lambda<-\delta,
\nonumber
\end{equation}
a singular integer power potential,
\begin{equation}
V(r)= \frac{\lambda}{r} +\frac{\mu}{r^2}+\frac{\chi}{r^3}+\frac{\tau}{r^4},
\nonumber
\end{equation}
and a singular anharmonic potential
\begin{equation}
V(r)= \omega r^2+ \frac{\epsilon}{r^2} +\frac{\sigma}{r^4}+\frac{\chi}{r^6},
\nonumber
\end{equation}
where all quantities different from 
independent variable $r$ are various potential parameters \cite{PBss}.

In the case of both Xie and Chen et al. 
modified Manning potentials with three parameters we have succeeded in determining 
{\em odd} parity eigenstates. Note that the original Ansatz
(\ref{XiA}) by Xie \cite{Xi} and the Ansatz (\ref{CMpA}) of Chen et al. \cite{CWX} can 
capture only {\em even} parity solutions. The odd parity solutions can be obtained by replacing 
$\phi(x)$ in the Ansatz (\ref{XiA}) by $\tanh x\, \phi(x)$, and by modifying the Ansatz (\ref{CMpA}) 
of Chen et al. \cite{CWX} to (\ref{CMpAe}) by adding an extra $\sinh x$ factor.
(Computational details have been relegated to the online supplementary material Secs. \ref{sc:3pXMmm} and \ref{sc:3pCMmm}.)
Parity resolved solution for the DSHG system can be obtained by going from the 
Ansatz (\ref{dshgA}) to the Ansatz (\ref{hRAd}).

For both the hyperbolic Razavy potential of Sec. \ref{sc:hRp} and the perturbed double sinh-Gordon system
of Sec. \ref{sc:pdshg} either substitution of independent variable $z=\cosh^2x$ or $z=\sinh^2 x$ 
is possible to transform the Schr\"odinger equation into (\ref{qese}).
That is illustrated in the online supplementary material Sec. \ref{sec:zshs}.

\subsection{The condition of $sl_2$ algebraization and an algebraic Heun operator}
\label{sec:dissl2}
As alluded to earlier, the {\em baseline} condition (\ref{bnc}) reappears in the 
functional Bethe Ansatz method (cf. Theorem 4 and Remark 9 of Ref. \cite{AMcp}; 
Eqs. (1.8-10) of Ref. \cite{Zh}), or as one of the conditions of $sl_2$ algebraization \cite{T16,AMcp,Zh6} - see
e.g. the condition $2\nu(2\nu-1)a_3+2\nu b_2+c_1=0$ for the $sl(2,\mathbb{R})$ spin $\nu$ representation of the
Heun operator of Turbiner \cite[Eq. (6)]{T16}, 
\begin{equation}
 H_e =(a_3 z^3 + a_2 z^2 + a_1 z) d_z^2 + (b_2 z^2 + b_1 z + b_0) d_z + c_1z,
\label{heun}
\end{equation}
when recast in our notation [cf. Eq. (\ref{qese})]. The operator (\ref{heun}) is defined up to additive constant $c_0$ -- 
it is the reference point for the spectral parameter and coincides with the accessory parameter in the Heun equation
\cite{T16}. When the baseline condition is satisfied, $H_e$ can be recast
in terms of the generators $J$'s of the $sl(2,R)$-Lie algebra \cite[Eq. (2)]{T16}
\begin{equation}
H = t^{+0} J_+ J_0 + t^{+-} J_+ J_- + t^{00} J_0 J_0 + t^{0-} J_0 J_- + B^+ J_+ + B^0 J_0 + B^- J_-,
\label{top}
\end{equation}
where $t^{+0}, t^{+-}, t^{00}, t^{0-}$ and $B^+, B^0, B^-$ are constants, with the correspondence
\begin{equation}
a_3=t^{+0},\quad b_2=t^{+0} (1 - 3\nu)+B^+,\quad c_1=2\nu(\nu t^{+0}-B^+).
\label{connect}
\end{equation}

To each two different points of the baseline there correspond two {\em different} algebraic Heun operators,
simply because they are determined by different constants $t^{0+}$, $B^+$ (\ref{top})
in the expansion in terms of the generators $J$'s of the $sl(2,R)$-Lie algebra.
On the $n$-th baseline energy $E$, and hence also the parameter $c_1$, even if it were formally dependent 
on energy, remain constant for the coupling constant metamorphosis QES examples 
of Sec. \ref{sc:nwop}. In particular, we have
\begin{equation}
c_1=-4n \sqrt{V_1}
\label{c1Mn}
\end{equation}
for the modified Manning potential with three parameters for both even [cf. Eqs. (\ref{XMec}), (\ref{epMe})]
and odd parity cases [cf. Eqs. (\ref{XMoc}), (\ref{opMe})].
For the Chen et al. modified Manning potential we have 
\begin{equation}
c_1 = - n^2- 2n(\lambda_1+\lambda_2), \qquad c_1 =- n^2- n(2\lambda_1+2\lambda_2)-\frac14
\nonumber
\end{equation}
in the respective even parity case [cf. Eqs. (\ref{CmMc}), (\ref{eCmMe})] and 
odd parity case [cf. Eqs. (\ref{CmMco}), (\ref{oCmMe})].
For an electron in Coulomb and magnetic fields and relative 
motion of two electrons in an external oscillator potential
\begin{equation}
c_1=\epsilon = \alpha/\omega -(g+1/2).
\nonumber
\end{equation}
On the other hand, spectral parameter $c_0$ 
depends on one of the other model parameters.

An illustration of what happens in search of the exceptional spectrum of various Rabi models \cite{AMcp,Ks,Jd}
can be provided by the modified Manning potential with three parameters of Sec. \ref{sc:3pM}
by selecting $\sqrt{V_1}$ as an independent spectral variable.
Any change of $\sqrt{V_1}$ induces a translation on the corresponding 
baseline in both the even [cf. Eq. (\ref{epMe})] and odd [cf. Eq. (\ref{opMe})]
parity cases. During those translations, the value of $c_1$ varies according to (\ref{c1Mn})
and the value of $b_2$ changes according to Eqs. (\ref{XMec}), (\ref{XMoc}).
Because of (\ref{heun}), (\ref{connect}), each different value of $b_2$, or $c_1$, corresponds to a {\em different}
$sl_2$ operator $H$ in (\ref{top}).

\subsection{${\cal P}(n)$ has only {\em real} and {\em simple} roots}
\label{sec:discsz}
Let us first introduce $p_{nk}$, $1\le k\le n+1$, through
\begin{equation}
 P_{nk}= \frac{p_{nk}}{\prod_{l=1}^k F_1(n-l)},\qquad - {\cal P}(n) = \frac{p_{n,n+1}}{\prod_{l=1}^n F_1(n-l)},
\nonumber
\end{equation}
while reminding that $F_1(n-l)\ne 0$ has been assumed for $1\le l\le n$. 
Now our original TTRR (\ref{gm0rcp}), together with the definition of the 
constraint polynomial (\ref{spc}), can be recast as a TTRR
\begin{equation}
p_{nk}= - F_{0} (n+1-k) p_{n,k-1} - F_{-1}(n+2-k) F_1(n+1-k) p_{n,k-2},\qquad 1\le k\le n+1,
\label{gm0rcpf}
\end{equation}
with the initial condition $p_{n,-1}=0$, $p_{n0}=1$.

In what follows we compare (\ref{gm0rcpf}) against the canonical TTRR for monic polynomials,
\begin{equation}
P_k(x)=(x-d_k)P_{k-1}(x)-\lambda_k P_{k-2}(x),\qquad k\ge 1,
\label{bttrr}
\end{equation}
with the initial condition $P_{-1}=0$, $P_{0}=1$. 
For any given $\{\lambda_k\}_{k=2},\, \{d_k\}_{k=1}\in\mathbb{C}$ the TTRR (\ref{bttrr}) generates 
an orthogonal polynomial system (OPS) if and only if $\lambda_k\ne 0$, $k\ge 2$ 
(cf. Favard's theorem - e.g. Theorem 4.4 of Chihara's book \cite{Chi}). 
Moreover:
\begin{itemize}

\item[({\bf a})]
If $\{P_k(x)\}$ satisfies the TTRR (\ref{bttrr}) with $\lambda_k\ne 0$ for $2\le k\le N$,
then $P_k(x)$  and $P_{k-1}(x)$ cannot have a common zero for $k\le N$
\cite[Exercise 4.3]{Chi}. If they had a common zero $x_0$, then necessarily 
$P_{k-2}(x_0)=P_{k-3}(x_0)=\ldots =P_0(x_0)=0$. But that contravenes the initial condition
$P_0(x)\equiv 1$.

\item[({\bf b})]
A unique moment functional ${\cal L}$ is {\em positive} definite 
if and only if $d_k$ and $\lambda_k>0$ are real, 
and additionally $\lambda_k>0$ ($k\ge 1$) \cite[p. 22]{Chi}. In the latter case \cite[p. 22]{Chi}
\begin{equation}
{\cal L}[P_k^2(x)]=\prod_{j=1}^{k+1}\lambda_j>0,\qquad k\ge 0. 
\label{chn}
\end{equation}
Under the above conditions the zeros of $P_k(x)$ are 
(i) all {\em real} and {\em simple}, and 
(ii) located in the interior of the supporting set for ${\cal L}$ \cite[Theorem 5.2]{Chi}.
Obviously, if (i) holds for the zeros of $P_k(x)$, the same is true also for the zeros of $P_k(-x)$.
But the latter are generated with $x$ replaced by $-x$ in (\ref{bttrr}). 

\end{itemize}
We can associate TTRR (\ref{gm0rcpf}) with TTRR (\ref{bttrr})
by  identifying $\lambda_k=F_{-1}(n+2-k)F_1(n+1-k)$ for $2\le k\le n+1$.
The baseline condition $F_1(n)=0$ implies $\lambda_1=0$. 
Because such a $\lambda_1$ multiplies $p_{n,-1}\equiv 0$ in (\ref{gm0rcpf})
nothing changes there if one assumes formally $\lambda_1\ne 0$.
Indeed, once the initial condition $P_{-1}=0$ is imposed (\ref{bttrr}) one has a freedom 
to select $\lambda_1$ according to one needs.
One finds that the following applies for the TTRR (\ref{gm0rcpf}) for $2\le k\le n+1$:
\begin{enumerate}

\item Xie \cite{Xi} modified Manning potential: TTRR (\ref{gm0rcpf}) is 
equivalent to TTRR (\ref{bttrr}) with
$x=V_3$, $\lambda_k=F_{-1}(n+2-k)F_1(n+1-k)>0$ in (\ref{3mF}), (\ref{3mFo});

\item Chen et al. \cite{CWX} modified Manning potential: 
 TTRR (\ref{gm0rcpf}) is equivalent to TTRR (\ref{bttrr}) with 
$x=-V_2/(4g)$, $\lambda_k=F_{-1}(n+2-k)F_1(n+1-k)>0$, provided that $g>0$ in (\ref{3mFv}), (\ref{3mFvo});

\item an electron in Coulomb and magnetic fields: 
 TTRR (\ref{gm0rcpf}) is equivalent to TTRR (\ref{bttrr}) with
$x=-\beta$, $\lambda_k=F_{-1}(n+2-k)F_1(n+1-k)>0$, provided that $g>0$ in (\ref{3CF});

\item the hyperbolic Razavy potential: 
TTRR (\ref{gm0rcpf}) is equivalent to TTRR (\ref{bttrr}) with 
$x=-E$, $\lambda_k=F_{-1}(n+2-k)F_1(n+1-k)<0$, provided that $\alpha>-1/2$, $\xi>0$ in (\ref{RzvF});

\item the double sinh-Gordon system (DSHG): 
TTRR (\ref{gm0rcpf}) is equivalent to TTRR (\ref{bttrr}) with 
$x=-E$, $\lambda_k=F_{-1}(n+2-k)F_1(n+1-k)>0$, provided that $\xi>0$ in (\ref{DSFpm});

\item the perturbed DSHG: 
TTRR (\ref{gm0rcpf}) is equivalent to TTRR (\ref{bttrr}) with 
$x=-E$, $\lambda_k=F_{-1}(n+2-k)F_1(n+1-k)<0$, provided that $\beta>-1/2$ and $\xi>0$ in (\ref{KMFp}).

\end{enumerate}
Therefore, for all the cases considered here the TTRR (\ref{gm0rcpf}) 
defines a finite OPS $\{p_{nk},\, k=0,1,2,\ldots,n+1\}$ satisfying at least the
condition ({\bf a}). Furthermore, in the 1st to 3rd and 5th case the above property ({\bf b}) is also satisfied.
Thus in those cases each polynomial of the finite OPS $\{p_{nk},\, k=0,1,2,\ldots,n+1\}$,
and correspondingly $\{P_{nk},\, k=0,1,2,\ldots,n,\, {\cal P}(n)\}$,
is guaranteed to have only {\em real} and {\em simple} roots in a corresponding independent variable $x$
for any $n$th baseline. Even if the above root property need not to hold in general 
in the 4th and 6th case of the hyperbolic Razavy potential of Sec. \ref{sc:hRp}
and a perturbed double sinh-Gordon system of Sec. \ref{sc:pdshg}, respectively,  
we could still observe it for the parameters considered.

\subsection{${\cal P}(n)$ vs weak orthogonal polynomials of Lancosz-Haydock and Bender-Dunne}
\label{sec:discwo}
If some $\lambda_N=0$ in the TTRR (\ref{bttrr}), then one speaks about the so-called
{\em weak} orthogonal polynomials \cite[p. 23]{Chi}. Examples of weak orthogonal polynomials 
are provided by the Lanczos-Haydock finite-chains of polynomials \cite{Hd,Lnz}, 
later rediscovered by Bender and Dunne \cite{BD,AMhd}. In the above cases a corresponding TTRR
(\ref{bttrr}) determines the coefficients of a sought polynomial solution (\ref{snex}) 
beginning from that of its {\em lowest} degree upwards, reflected 
by the initial conditions on the two coefficients of the {\em lowest} degree
(cf. Eq. (5) of Ref. \cite{BD})
\begin{eqnarray}
P_{nn}=P_0 = 1 \qquad {\rm and} \qquad P_{n,n-1}=P_1(E) = E.
\label{bde7}
\end{eqnarray}
Contrary to that, a corresponding TTRR
(\ref{bttrr}) in our case determines the coefficients of a sought polynomial solution (\ref{snex}) 
beginning from that of its {\em highest} degree downwards, which is reflected by
the initial condition $P_{n0}=P_n = 1$,
i.e. involving the coefficient of the {\em highest} degree of 
a sought polynomial solution (\ref{snex}).
This bring us to two important differences relative to the weak orthogonal Bender-Dunne polynomials:
\begin{itemize}

\item[(i)] First, we cannot guarantee in our case that the conditions (\ref{bde7}) will be satisfied, 
simply because our TTRR (\ref{gm0rcp}), (\ref{spc}), or (\ref{gm0rcpf}), 
run in the {\em opposite} direction. Consequently, one may well end up with, 
and cannot exclude that, e.g. $P_{nn}=P_0 = 0$.

\item[(ii)] Second, with $\lambda_N=0$ in the TTRR (\ref{bttrr}), 
the quasi-exact energy eigenvalues
are the roots of a critical polynomial $P_N$ of a corresponding weak orthogonal polynomial 
sequence that is determining $N$ energy levels in the $N$-dimensional 
polynomial subspace $\{1,z,\, z^2,\ldots,\, z^{N-1}\}$ \cite{Hd,Lnz,BD,AMhd}. 
Hence the polynomial degree of solutions need not to be $N$.
Yet in our case all the polynomial solution on the $n$-baseline are of $n$th degree 
by construction \cite{AMcp}. Therefore, our constraint polynomials ${\cal P}(n)$
are {\em not} necessarily the critical polynomials of the weak orthogonal Bender-Dunne polynomials.

\end{itemize}

A TTRR may possess a unique minimal (or dominated) solution \cite{Gt,Ht}.
It is interesting to recall that in the case when only the minimal solutions are the required 
physical solutions \cite{AMef}, then the whole physical spectrum of the model 
(i.e. including non-algebraic part of the spectrum) coincides 
with the support $\mathfrak{S}$ of a positive-definite moment functional ${\cal L}$ of 
corresponding discrete orthogonal polynomials \cite{AMef}.
Therefore not only the algebraic part of the spectrum may be closely related to orthogonal polynomials.

\section{Conclusions}
\label{sec:conc}
Recently developed general constraint polynomial approach was shown to replace
a set of algebraic equations of the functional Bethe Ansatz method by a single polynomial constraint.
As the proof of principle, the usefulness of the method has been demonstrated 
for a number of quasi-exactly solvable potentials of the Schr\"odinger equation, 
enabling one to straightforwardly determine eigenvalues and wave functions.

Our constraint polynomials, which were shown to be different
from the weak orthogonal Bender-Dunne polynomials, appear to be yet another class of polynomials closely related 
to the spectrum of quasi-exactly solvable models. For the models considered here, constraint polynomials
terminated a finite chain of orthogonal polynomials characterized by 
a positive-definite moment functional ${\cal L}$, implying that a corresponding
constraint polynomial has only real and simple zeros.

\section{Acknowledgments}
\label{sec:ack}
AM acknowledges discussions with A. M. Ishkhanyan, B. M. Rodr\'{\i}guez-Lara, and M. Znojil
in different stages of this work. The work of AEM was supported by the Australian 
Research Council and UNSW Scientia Fellowship.

\newpage


\newpage

\begin{center}
Online supplementary material
\end{center}

\newpage

\section{Generic coordinate transformation}
\label{sc:gtr}
\begin{equation}
\psi(x)=Q(z)\phi(z),\qquad d_z Q(z)=K(z)Q(z),\qquad z=f(x),
\nonumber
\end{equation}
implies $d_z^2 Q(z)=[K'(z)+K^2(z)]Q(z)$ and 
\begin{eqnarray}
& d_z [Q(z)\phi(z)]=Q(z)[\phi'(z)+K(z)\phi(z)],&
\nonumber\\
&
d_z^2 [Q(z)\phi(z)]=Q(z)\left\{ \phi''(z)+2K(z)\phi'(z)+[K'(z)+K^2(z)]\phi\right\},&
\nonumber\\
& d_x=f'(x)d_z,\qquad d_x^2=[f'(x)]^2 d_z^2+f''(x)d_z.&
\end{eqnarray}
The Schr\"odinger equation (\ref{SE}) then becomes
\begin{eqnarray}
\lefteqn{
[f'(x)]^2 \phi''(z)+\left\{ 2[f'(x)]^2K(z)+f''(x)\right\}\phi'(z)}
\nonumber\\
&& \, +\left\{E-V+[f'(x)]^2[K'(z)+K^2(z)]+f''(x)K(z)\right\}\phi(z)=0.
\label{SEt}
\end{eqnarray}

As a slight variation of (\ref{SEt}) we have with $z=f(x)$ for
\begin{equation}
\psi(x)=Q(x)\phi(z),\qquad d_x Q(x)=K(x)Q(x),\qquad d_x^2 Q(x)=[K'(x)+K^2(x)]Q(x),
\nonumber
\end{equation}
\begin{eqnarray}
& d_x=f'(x)d_z,\qquad d_x^2=[f'(x)]^2 d_z^2+f''(x)d_z,&
\nonumber\\
& d_x Q(x) d_x \phi(z)=Q(x)[K(x) f'(x) \phi'(z)],&
\nonumber\\
&
d_x^2 [Q(x)\phi(z)]=Q(x)\left\{ [f'(x)]^2 \phi''(z)+f''(x) \phi'(z)+2K(x)f'(x) \phi'(z) 
\right. &
\nonumber\\
&
\left.
+ [K'(x)+K^2(x)]\phi(z) \right\},&
\nonumber
\end{eqnarray}
and
\begin{equation}
[f'(x)]^2 \phi''(z)+\left[ 2f'(x) K(x)+f''(x)\right] \phi'(z) +\left[E-V+ K'(x)+K^2(x) \right]\phi(z)=0.
\label{SEtc}
\end{equation}

\subsection{Xie modified Manning potential with 
three parameters and $z=\tanh^2 x$}
\label{sc:3pXMmm}
In the case of the Ansatz (\ref{XiA}) for the Xie modified Manning potential (\ref{XmMp}) 
with three parameters of Sec. \ref{sc:3pM},
\begin{eqnarray}
K(z) &=& \frac{\sqrt{V_1}}{2}-\frac{\sqrt{-E}}{2(1-z)},\qquad
K'(z) = -\frac{\sqrt{-E}}{2(1-z)^2},
\nonumber\\
K'(z)+K^2(z)&=& \frac{V_1}{4} -\frac{\sqrt{V_1} \sqrt{-E}}{2(1-z)}-\frac{E+2\sqrt{-E}}{4(1-z)^2},
\nonumber\\
f'(x)&=& 2\tanh x \mbox{ sech}^2 x,
\nonumber\\
{}
[f'(x)]^2 &=& 4z(1-z)^2,\qquad f''(x)=2(1-z)^2 -4z(1-z)=(1-z)(2-6z),
\nonumber\\
2[f'(x)]^2K(z)+f''(x) &=&(1-z)\left[4z(1-z)\sqrt{V_1}-4z\sqrt{-E}-6z+2\right].
\nonumber
\end{eqnarray}
Hence from (\ref{SEt})
\begin{eqnarray}
A(z)&=& \frac{1}{1-z}[f'(x)]^2=4z(1-z),
\nonumber\\
B(z)&=& \frac{1}{1-z}\left\{ 2[f'(x)]^2 K(z)+f''(x)\right\}=4z(1-z)\sqrt{V_1}-4z\sqrt{-E}-6z+2.
\nonumber
\end{eqnarray}

Given that
\begin{eqnarray}
&[f'(x)]^2 [K'(z) + K^2(z)] = (1-z) \left[V_1 z(1-z) - 2 z\sqrt{V_1} \sqrt{-E} -\frac{E+2\sqrt{-E}}{1-z}\, z\right],&
\nonumber\\
&f''(x)K(z) = \sqrt{V_1} (1-z)(1-3z)-\sqrt{-E}(1-3z),&
\nonumber\\
& \frac{E}{1-z} -\frac{E+2\sqrt{-E}}{1-z}\, z -\frac{\sqrt{-E}}{1-z}\, (1-3z) = E-\sqrt{-E},&
\nonumber
\end{eqnarray}
we have eventually from (\ref{SEt})
\begin{eqnarray}
C(z)&=& \frac{1}{1-z}\left\{ E-V+[f'(x)]^2[K'(z)+K^2(z)]+f''(x)K(z)\right\}
\nonumber\\
&=& E-\sqrt{-E} +V_1(1-z)^2+V_2(1-z)+V_3+\sqrt{V_1} (1-3z)
\nonumber\\
&&
+V_1 z(1-z) - 2 z\sqrt{V_1} \sqrt{-E}
\nonumber\\
&=& E-\sqrt{-E} +V_1-V_1z+V_2(1-z)+V_3+\sqrt{V_1}(1-3z) - 2 z\sqrt{V_1} \sqrt{-E}
 \nonumber\\
&=& z(-V_1-V_2-3\sqrt{V_1}-2 \sqrt{V_1} \sqrt{-E}) + E-\sqrt{-E}
\nonumber\\
&& +V_1+V_2+V_3+\sqrt{V_1}.
\nonumber
\end{eqnarray}
One recovers the polynomial coefficients (\ref{XMec}) by multiplying the current $A(z),\, B(z),\, C(z)$
by minus one.

Provided that $\phi(x)$ in the Ansatz (\ref{XiA}) is replaced by $\tanh x\, \phi(x)$,
we have the following changes in the above formulas:
\begin{eqnarray}
K(z) &=& \frac{\sqrt{V_1}}{2}-\frac{\sqrt{-E}}{2(1-z)}+ \frac{1}{2z},\qquad
K'(z) = -\frac{\sqrt{-E}}{2(1-z)^2}-\frac{1}{2z^2},
\nonumber\\
\Delta [K'(z)+K^2(z)] &=& -\frac{1}{2z^2} + \frac{1}{4z^2}+ 
 \frac{1}{z} \left[\frac{\sqrt{V_1}}{2}-\frac{\sqrt{-E}}{2(1-z)}\right]
\nonumber\\
&=& -\frac{1}{4z^2} + \frac{\sqrt{V_1}}{2z}-\frac{\sqrt{-E}}{2z(1-z)},
\nonumber\\
\Delta \left\{ 2[f'(x)]^2K(z)+f''(x) \right\} &=& 2[f'(x)]^2 \Delta K(z)=8z(1-z)^2 \frac{1}{2z}=4(1-z)^2.
\nonumber
\end{eqnarray}
In order to recover the polynomial coefficients (\ref{XMoc}) for the odd parity Ansatz of Sec. \ref{sc:3pMo}
it suffices to focus only on the above changes indicated by $\Delta$.
One finds immediately
\begin{eqnarray}
A(z)&=& \frac{1}{1-z}[f'(x)]^2=4z(1-z),
\nonumber\\
B(z)&=& 4z(1-z)\sqrt{V_1}-4z\sqrt{-E}-6z+2+ 4(1-z)
\nonumber\\
&=& - 4z^2\sqrt{V_1}-z(4\sqrt{-E}-4\sqrt{V_1}+10)+6.
\nonumber
\end{eqnarray}
Given that
\begin{eqnarray}
&\Delta[f'(x)]^2 [K'(z) + K^2(z)] =(1-z)^2 \left(-\frac{1}{z} + 2 \sqrt{V_1} -\frac{2\sqrt{-E}}{1-z}\right),&
\nonumber\\
&\Delta [f''(x)K(z)] =(1-z)\, \frac{1-3z}{z} ,&
\nonumber\\
& \Delta C(z)=\frac{1-3z}{z}-\frac{1-z}{z} + 2(1-z) \sqrt{V_1} -2\sqrt{-E}
\nonumber\\
&=-2z \sqrt{V_1} +2\sqrt{V_1} -2\sqrt{-E}-2.&
\nonumber
\end{eqnarray}
One recovers the polynomial coefficients (\ref{XMoc}) after 
multiplication of the current $A(z),\, B(z),\, C(z)$ by minus one.

\subsection{Chen et al. modified Manning potential with 
three parameters and $z=-\sinh^2 x$}
\label{sc:3pCMmm}
For the Ansatz (\ref{CMpA}) in the case of the Chen et al. 
modified Manning potential (\ref{CmMp}) with three parameters of Sec. \ref{sc:3pM}
on arrives at (\ref{SEtc}). Now with $z=f(x)=-\sinh^2 x$ and the Ansatz (\ref{CMpAe}),
\begin{equation}
f'(x)= -\sinh 2x, \qquad [f'(x)]^2 = \sinh^2 2x= 4 \sinh^2 x\cosh^2 x= 4z(z-1),
\nonumber
\end{equation}
\begin{eqnarray}
K(x) &=& 2\lambda_1\tanh x +\frac{\lambda_2 g \sinh 2x}{1+g\cosh^2 x}+\coth x,
\nonumber\\
K'(x) &=& \frac{2\lambda_1}{\cosh^2 x} +\frac{2\lambda_2 g \cosh 2x}{1+g\cosh^2 x}
- \frac{\lambda_2 g^2 \sinh^2 2x}{(1+g\cosh^2 x)^2} - \frac{1}{\sinh^2 x},
\nonumber\\
K^2(x) &=& \left(2\lambda_1\tanh x +\frac{\lambda_2 g \sinh 2x}{1+g\cosh^2 x}+\coth x\right)^2,
\nonumber\\
\Delta[K'(z)+K^2(z)]&=&- \frac{1}{\sinh^2 x} + \coth^2 x +4\lambda_1 +\frac{4\lambda_2 g \cosh^2 x}{1+g\cosh^2 x}
\nonumber\\
&=& 4\lambda_1 +1 +\frac{4\lambda_2 g \cosh^2 x}{1+g\cosh^2 x}=4\lambda_1+4\lambda_2+1 -\frac{4\lambda_2 }{1+g\cosh^2 x},
\nonumber\\
{}
\Delta[ 2f'(x) K(x)+f''(x)] &=& -2\sinh 2x \coth x = -4 \cosh^2 x=4(z-1).
\nonumber
\end{eqnarray}
Here and below $\Delta$ indicates the change of the term preceded by $\Delta$ 
obtained from the Ansatz (\ref{CMpAe}) relative to that resulting from the Ansatz (\ref{CMpA}).

On multiplying (\ref{SEtc}) by $1+g\cosh^2 x=1+g(1-z)$ one finds 
the polynomial coefficient of $\phi''(z)$,
\begin{eqnarray}
4z(z-1)[1+g(1-z)]&=& -4z[g z^2-z(1+2g)+1+g]
\nonumber\\
&=&-4g\left[z^3-z^2\left(2+\frac1g\right)+1+\frac1g \right].
\nonumber
\end{eqnarray}
One can reproduce 
the polynomial coefficient $A(z)$ of $\phi''(z)$ in Eqs. (\ref{CmMc}) after factoring out the prefactor $-4g$.
Similarly one determines $\Delta B(z)$ from
\begin{eqnarray}
\Delta B(z) &=& -\frac{1}{4g}\, \Delta[ 2f'(x) K(x)+f''(x)] [1+g(1-z)]
\nonumber\\
&=& \frac{1}{g}\, (1-z) [1+g(1-z)]= z^2 -z\, \frac{1+2g}{g}+\frac{1+g}{g},
\label{DtB}
\end{eqnarray}
and $\Delta C(z)$ from
\begin{eqnarray}
\Delta C(z) &=& -\frac{1}{4g}\, \Delta[K'(z)+K^2(z)] [1+g(1-z)]
\nonumber\\
&=& -\frac{1}{4g}\,\left[ 4\lambda_1+4\lambda_2+1 -\frac{4\lambda_2 }{1+g(1-z)}\right] [1+g(1-z)]
\nonumber\\
&=& \frac{1}{4g}\,\left\{ 4\lambda_2 - (4\lambda_1+4\lambda_2+1) [1+g(1-z)]\right\} 
\nonumber\\
&=& z \left(\lambda_1+\lambda_2+\frac14\right) 
 - \left(\lambda_1+\lambda_2+\frac14\right)\frac{1+g}{g}+ \frac{\lambda_2}{g}\cdot
\label{DtC}
\end{eqnarray}

\subsection{Hyperbolic Razavy potential}
\label{sc:hRza}
$z=\cosh^2x$ implies
\begin{eqnarray}
d_x &=& 2 \cosh x \sinh x\, d_z = \sinh 2x\, d_z,
\nonumber\\
d_x^2 &=& d_x(\sinh 2x\, d_z) = 2 \cosh 2x\, d_z+ \sinh^2 2x\, d_z^2
= 2 \cosh 2x\, d_z+ 4 (\sinh^2 x \cosh^2 x)\, d_z^2,
\nonumber\\
&=& 2(2z-1)\, d_z+ 4 z(z-1)\, d_z^2,
\nonumber\\
\sinh 2x\, d_x &=& \sinh^2 2x\, d_z=4 \sinh^2 x \cosh^2x \, d_z= 4z(z-1)\, d_z.
\label{chss}
\end{eqnarray}
For the hyperbolic Razavy potential (\ref{hRp}), and with $\xi\to \xi/2$ in the expression for $Q(x)$ above,
one finds
\begin{eqnarray}
&&
-\frac{\xi^2}{4}\, \sinh^2 2x + (N+1) \xi \cosh(2x)+\frac{\xi^2}{4}\,\sinh^2 2x +(\alpha+\beta)^2
\nonumber\\
&& -2 \xi(\alpha\sinh^2 x+\beta \cosh^2x)
- \xi \cosh(2x)
\nonumber\\
&&
= N \xi \cosh(2x) +(\alpha+\beta)^2 -2 \xi(\alpha\sinh^2 x+\beta \cosh^2x),
\label{Qe}
\end{eqnarray}
and
\begin{eqnarray}
\lefteqn{
\left[d_x^2 -\frac{\xi^2}{4}\, \sinh^2 2x + (N+1)\xi \cosh(2x)\right] (Q\phi) =
}
\nonumber\\
&&
Q\left[ \vphantom{\frac12} d_x^2 +
 \left(-\xi \sinh 2x+2\alpha \tanh x+2\beta \coth x \right)d_x
\right.
\nonumber\\
&&
\left.
 +E+\, N \xi \cosh(2x) +(\alpha+\beta)^2 -2 \xi(\alpha\sinh^2 x+\beta \cosh^2x)
\right]\phi.
\label{QetahR}
\end{eqnarray}
Eventually one makes use of (\ref{chss}) to deduce that
\begin{eqnarray}
\lefteqn{
d_x^2 + \left(-\xi \sinh 2x+2\alpha \tanh x+2\beta \coth x \right)d_x
=
}
\nonumber\\
&&4z(z-1)d_z^2 +\left[2(2z-1) -4\xi z(z-1)+ 4\alpha(z-1) +4\beta z\right]d_z.
\label{Rzvddz}
\end{eqnarray}

\subsection{DSHG}
\label{sc:dshga}
For the Ansatz (\ref{dshgA}) we have with $z=e^{2x}$
\begin{eqnarray}
& K(z) = \frac{1-M}{2z} -\frac{\xi}{4z} \left(z-\frac{1}{z} \right),
\qquad
K'(z) = -\frac{1-M}{2z^2} -\frac{\xi}{2z^3},&
\nonumber\\
&[f'(x)]^2=4z^2,\qquad f''(x)=4z,\qquad 
V(z)=\left[\frac{\xi}{2} \left(z+\frac{1}{z} \right) -M\right]^2.&
\nonumber
\end{eqnarray}
Hence from (\ref{SEt})
\begin{eqnarray}
A(z)&=& [f'(x)]^2=4z^2,
\nonumber\\
B(z)&=& 2[f'(x)]^2 K(z)+f''(x)=4z \left[1-M -\frac{\xi}{2}\left(z-\frac{1}{z}\right)\right]+4z
\nonumber\\
&=& - 2\xi z^2 + 4z(2-M)+2\xi.
\nonumber
\end{eqnarray}

Given that
\begin{eqnarray}
[f'(x)]^2 K^2(z)-V(z) &=& \left[1-M -\frac{\xi}{2} \left(z-\frac{1}{z} \right)\right]^2
 -\left[\frac{\xi}{2} \left(z+\frac{1}{z} \right) -M\right]^2
\nonumber\\
 &=& 1+ 2M\xi z -2M-\xi \left(z-\frac{1}{z} \right) -\xi^2,
\nonumber
\end{eqnarray}
\begin{equation}
f''(x)K(z)=4zK(z)=2(1-M) -\xi \left(z-\frac{1}{z} \right),
\nonumber
\end{equation}
\begin{equation}
[f'(x)]^2 K'(z)=-2(1-M) -\frac{2\xi}{z},
\nonumber
\end{equation}
we have eventually from (\ref{SEt})
\begin{eqnarray}
C(z)&=& E-V+[f'(x)]^2[K'(z)+K^2(z)]+f''(x)K(z)
\nonumber\\
&=& 2\xi(M-1) z + E+ 1 -2M -\xi^2.
\nonumber
\end{eqnarray}

\subsection{Perturbed DSHG}
\label{sc:pdshga}
\begin{eqnarray}
Q &:=& e^{-\frac{\xi}{2}\,\cosh 2x}\, (\cosh x)^\alpha (\sinh x)^\beta,
\nonumber\\
Q' &:=& \left(-\xi \sinh 2x+\alpha \tanh x+\beta \coth x \right) \, Q,
\nonumber\\
Q'' &:=& Q\left[
\left(-\xi \sinh 2x+\alpha \tanh x+\beta \coth x \right)^2 -2 \xi \cosh 2x
 + \frac{\alpha}{\cosh^2 x}-\frac{\beta}{\sinh^2 x} \right],
 \nonumber\\
\lefteqn{
\left(-\xi \sinh 2x+\alpha \tanh x+\beta \coth x \right)^2 =
}
\nonumber\\
&&
\xi^2 \sinh^2 2x+\alpha^2 \tanh^2 x+\beta^2 \coth^2 x
-4 \xi\alpha \sinh^2 x -4 \xi\beta\cosh^2 x + 2\alpha\beta, 
\nonumber\\
&&
\alpha^2 \tanh^2 x+ \frac{\alpha}{\cosh^2 x} =\alpha^2 - \frac{\alpha(\alpha-1)}{\cosh^2 x},
\nonumber\\
&&
\beta^2 \coth^2 x -\frac{\beta}{\sinh^2 x}=\beta^2 + \frac{\beta(\beta-1)}{\sinh^2 x},
\nonumber\\
&&
- (\xi \cosh 2x -M)^2+ 
\xi^2 \sinh^2 2x -4 \xi\alpha \sinh^2 x -4 \xi\beta\cosh^2 x +(\alpha+\beta)^2-2 \xi \cosh 2x
\nonumber\\
&&
= -\xi^2 +2\xi(2z-1)M-M^2+(\alpha+\beta)^2 -4 \xi\alpha(z-1) -4 \xi\beta z-2 \xi(2z-1)
\nonumber\\
&&
= -\xi^2 -M^2+(\alpha+\beta)^2 + 2 \xi(2\alpha-M+1) +4 \xi z(M-\alpha-\beta-1), 
\nonumber
\end{eqnarray}
where $z=\cosh^2 x$. Hence
\begin{eqnarray}
\lefteqn{
\left[d_x^2 - (\xi \cosh 2x -M)^2\right] (Q\phi) =
}
\nonumber\\
&&
Q\left[ \vphantom{\frac12} d_x^2 + 2\left(-\xi \sinh 2x+\alpha \tanh x+\beta \coth x \right)d_x
\right.
\nonumber\\
&&
\left.
 -M^2-\xi^2+(\alpha+\beta)^2 + 2 \xi(2\alpha-M+1) +4 \xi z(M-\alpha-\beta-1)
\right.
\nonumber\\
&&
\left.
- \frac{\alpha(\alpha-1)}{\cosh^2 x}
+ \frac{\beta(\beta-1)}{\sinh^2 x} \right]\, \phi.
\label{Qeta}
\end{eqnarray}
Eventually one makes use of (\ref{chss}) to deduce that
\begin{eqnarray}
\lefteqn{
d_x^2 + 2\left(-\xi \sinh 2x+\alpha \tanh x+\beta \coth x \right)d_x
=
}
\nonumber\\
&&4z(z-1)d_z^2 +\left[2(2z-1) -8\xi z(z-1)+ 4\alpha(z-1) +4\beta z\right]d_z.
\nonumber 
\end{eqnarray}
The latter differs from (\ref{Rzvddz}) by the substitution $\xi \to 2\xi$.

\section{Independent variable $z=\sinh^2 x$}
\label{sec:zshs}
For both the hyperbolic Razavy potential of Sec. \ref{sc:hRp} and the perturbed double sinh-Gordon system
of Sec. \ref{sc:pdshg} either substitution of independent variable $z=\cosh^2x$ or $z=\sinh^2 x$ 
is possible to transform the Schr\"odinger equation into (\ref{qese}).
The former substitution was used in the main text. Here we illustrate the possibility of the latter.
The substitution of independent variable $z=\sinh^2x$ implies on recalling
elementary formulas
\begin{eqnarray}
& 2 \cosh x \sinh x = \sinh 2x,\qquad \cosh 2x = 2 \sinh^2 x +1,&
\nonumber\\
&\sinh^2 2x= 4 \cosh^2 x \sinh^2 x,&
\nonumber
\end{eqnarray}
\begin{eqnarray}
d_x &=& 2 \cosh x \sinh x\, d_z = \sinh 2x\, d_z,
\nonumber\\
d_x^2 &=& 2 \cosh 2x\, d_z+ 4 (\sinh^2 x \cosh^2 x) d_z^2=2(2z+1)\, d_z+ 4 z(z+1)\, d_z^2,
\nonumber\\
\sinh 2x\, d_x &=& 4 \sinh^2 x \cosh^2x \, d_z= 4z(z+1)\, d_z.
\label{shss}
\end{eqnarray}

For the hyperbolic Razavy potential of Sec. \ref{sc:hRp} the neglected possibility of the 
substitution $z=\sinh^2 x$ implies in virtue of (\ref{shss}) that the Schr\"odinger equation is transformed into
\begin{eqnarray}
& 4z(z+1)\, d_z^2+\left[-4\xi z^2 +4(\alpha+\beta-\xi+1)z +2(2\beta+1)\right]\, d_z &
\nonumber\\
& + \left[2\xi (N-\alpha-\beta)z +E+ (\alpha+\beta)^2 +\xi(N-2\beta) \right], & 
\nonumber
\end{eqnarray}
which is (\ref{qese}) with
\begin{eqnarray}
& a_2 = 4,\qquad a_1 =4,& 
\nonumber\\
& b_2 = -4\xi,\qquad b_1 = 4(\alpha+\beta-\xi+1) ,\qquad b_0 = 2(2\beta+1),& 
\nonumber\\
& c_1 = 2\xi(N-\alpha-\beta),\qquad c_0 = E+(\alpha+\beta)^2+\xi(N-2\beta).& 
\nonumber
\end{eqnarray}
The necessary condition $F_1(n)=-4n\xi+2\xi(N-\alpha-\beta)=0$ remains the same 
and is solved as before by $N = 2n+\alpha+\beta$. On the $n$th baseline 
one has a slightly modified versions of (\ref{RzvF}) and (\ref{c0rzv}),
\begin{eqnarray}
& F_1(k)=4\xi (n-k), \qquad F_0(k)=4k (k+\alpha+\beta -\xi) +c_0(n), &
\nonumber\\
& F_{-1}(k)=2k(2k-1+2\beta),& 
\label{RzvFm}
\end{eqnarray}
where $c_0(n)= E+(\alpha+\beta)^2+\xi(n+\alpha-\beta)$.
Being a linear function, $F_1(k)$ in Eqs. (\ref{RzvF}), (\ref{RzvFm}) has for each $n$ only single zero. 
Hence the conditions (\ref{th1nsc})
are satisfied and there can always be only a unique polynomial solution for a given fixed set of parameters.

For the perturbed double sinh-Gordon system
of Sec. \ref{sc:pdshg}, the substitution $z=\sinh^2 x$ transforms Eq. (\ref{etapp}) 
in virtue of (\ref{shss}) into (\ref{qese}) with
\begin{eqnarray}
& a_2 = 4,\qquad a_1 = 4,& 
\nonumber\\
& b_2 =-8\xi,\qquad b_1 = 4(\alpha +\beta-2\xi+1),\qquad b_0 = 2(2\beta+1),& 
\nonumber\\
& c_1 = 4\xi(M-\alpha-\beta-1),\qquad
 c_0 = E -M^2- \xi^2+(\alpha+\beta)^2+2 \xi(M-2\beta-1).& 
\end{eqnarray}
The necessary condition $F_1(n)=-8n\xi+ 4 \xi(M-\alpha-\beta-1)=0$ remains the same as before 
and is solved by $M =2n+\alpha+\beta+1$. On the $n$th baseline one has in virtue of (\ref{grmchp})
\begin{eqnarray}
& F_1(k)=8\xi(n-k), \quad F_0(k)=4k(k+\alpha+\beta-2\xi+1)+c_0(n), &
\nonumber\\
& F_{-1}(k)=2k(2k-1+2\beta),& 
\label{KMFpm}
\end{eqnarray}
where
\begin{eqnarray}
c_0(n) &=& E-(2n+\alpha+\beta+1)^2- \xi^2+(\alpha+\beta)^2+2 \xi(\alpha-\beta+2n)
\nonumber\\
 &=& E -(2n+1)(2n+1+2\alpha+2\beta)- \xi^2+2 \xi(\alpha-\beta+2n)
 \nonumber
\end{eqnarray}
is, up, to a different sign of $2n$ in the last parenthesis, the same as in Eq. (\ref{c0dsh}).


\begin{thebibliography}{99}


\bibitem{HaS}
N. Hatami and M. R. Setare,
Exact solutions for a class of quasi-exactly solvable models: A unified treatment,
Eur. Phys. J. Plus {\bf 132}, 311 (2017).


\bibitem{PBss}
H. Panahi and M. Baradaran,
Unified treatment of a class of spherically symmetric potentials: quasi-exact solution,
arXiv:1607.04505.


\bibitem{BP}
M. Baradaran and H. Panahi,
Exact solutions of a class of double-well potentials: Algebraic Bethe ansatz,
arXiv:1712.06439.


\bibitem{BPdw}
M. Baradaran and H. Panahi,
Lie symmetry and the Bethe Ansatz solution of 
a new quasi-exactly solvable double-well potential,
Adv. High Energy Phys. 2017, 2181532
(arXiv:1702.06181 [math-ph]).


\bibitem{KK}
R. Koc and M. Koca,
A unified treatment of quasi-exactly solvable potentials I.,
arXiv:math-ph/0505002.


\bibitem{T16}
A. V. Turbiner,
The Heun operator as Hamiltonian,
J. Phys. A: Math. Theor. {\bf 49}, 26LT01 (2016)
(arXiv:1603.02053).
http://arxiv.org/abs/1603.02053


\bibitem{TPR16}
A. V. Turbiner,
One-dimensional quasi-exactly solvable Schr\"odinger equations, 
Phys. Rep. {\bf 642}, 1-71 (2016)
(arXiv:1603.02992).



\bibitem{Ishk18}
A. M. Ishkhanyan,
Schr\"odinger potentials solvable in terms of the general Heun functions,
Ann. Phys. {\bf 388}, 456-471 (2018)
(arXiv:1601.03360 [quant-ph]).


\bibitem{IshcH}
A. M. Ishkhanyan,
Schr\"odinger potentials solvable in terms of the confluent Heun functions,
Theoret. Math. Phys. {\bf 188}, 980-993 (2016).


\bibitem{Ishk17}
T. A. Ishkhanyan and A. M. Ishkhanyan,
Solutions of the bi-confluent Heun equation in terms of the Hermite functions,
Ann. Phys. {\bf 383} 79-91 (2017)
(arXiv:1608.02245 [quant-ph]).


\bibitem{AMcp}
A. Moroz, 
A unified treatment of polynomial solutions and constraint polynomials
of the Rabi models, J. Phys. A: Math. Theor. {\bf 51}, 295201 (2018)
(arXiv:1712.09371).


\bibitem{Xi}
Q.-T. Xie, 
New quasi-exactly solvable double-well potentials,
J. Phys. A: Math. Theor. {\bf 45}, 175302 (2012).


\bibitem{CWX}
B. Chen, Y. Wu, and Q. Xie, 
Heun functions and quasi-exactly solvable double-well potentials,
J. Phys. A: Math. Theor. {\bf 46}(3), 035301 (2013).


\bibitem{Mnn2}
M. F. Manning, 
Energy levels of a symmetrical double minima problem 
with applications to the NH${}_3$ and ND${}_3$ molecules,
J. Chem. Phys. {\bf 3}, 136-138 (1935).
 
 
\bibitem{Tr94}
A. Turbiner,
Two electrons in an external oscillator potential: The hidden algebraic structure,
Phys. Rev. A {\bf 50}, 5335-5337 (1994)
(arXiv:hep-th/9406018).


\bibitem{CH}
C.-M. Chiang and C.-L. Ho,
Charged particles in external fields as physical examples of 
quasi-exactly-solvable models: A unified treatment,
Phys. Rev. A {\bf 63}, 062105 (2001).


\bibitem{KM}
A. Khare and B.P. Mandal, 
New quasi-exactly solvable Hermitian as well 
as non-Hermitian PT-invariant potentials,
Pramana J. Phys. {\bf 73}, 387-395 (2009).


\bibitem{Rzv}
M. Razavy, 
An exactly soluble Schr\"odinger equation with a bistable potential,
Am. J. Phys. {\bf 48}, 285-288 (1980).


\bibitem{Zh}
Y.-Z. Zhang,
Exact polynomial solutions of second 
order differential equations and their applications,
J. Phys. A: Math. Theor. {\bf 45}, 065206 (2012)
(arXiv:1107.5090).


\bibitem{Zh6}
Y.-Z. Zhang,
Hidden $sl(2)$-algebraic structure in Rabi model and its 2-photon 
and two-mode generalizations,
Ann. Phys. (N.Y.) {\bf 375}, 460-470 (2016)
(arXiv:1608.05484).


\bibitem{Jd}
B. R. Judd,
Exact solutions to a class of Jahn-Teller systems,
J. Phys. C: Solid State Phys. {\bf 12}, 1685-1692 (1979).


\bibitem{Ks}
M. Kus, On the spectrum of a two-level system,
J. Math. Phys. {\bf 26}, 2792-2795 (1985).


\bibitem{Hd}
R. Haydock, 
``The recursive solution of the Schr\"{o}dinger equation," 
in: H. Ehrenreich, F. Seitz, D. Turnbull (Eds.),
Solid State Physics vol. 35, Academic Press, New York, 1980, pp. 215-294.


\bibitem{Lnz}
C. Lanczos, 
``An iteration method for the solution of the eigenvalue problem of linear 
differential and integral operators," 
J. Res. Nat. Bur. Stand. {\bf 45}, 255-282 (1950).


\bibitem{BD}
C. M. Bender and G. V. Dunne,
Quasi-exactly solvable systems and orthogonal polynomials,
J. Math. Phys. {\bf 37}, 6-11 (1996).


\bibitem{AMhd}
A. Moroz, 
Haydock's recursive solution of self-adjoint problems. Discrete spectrum,
Ann. Phys. (N.Y.) {\bf 351}, 960-974 (2014).


\bibitem{LL3}
L. D. Landau and E. M. Lifschitz, 
Quantum Mechanics (Pergamon, London 1960), ch. V, par. 35.


\bibitem{Zn0}
M. Znojil,
Comment on ``Conditionally exactly soluble class of quantum potentials''.
Phys. Rev. A {\bf 61}, 066101 (2000).
(arXiv:quant-ph/9811088).


\bibitem{FAS}
A. Faribault, O. El Araby, C. Str\"{a}ter, and V. Gritsev,
Gaudin models solver based on the correspondence 
between Bethe ansatz and ordinary differential equations,
Phys. Rev. B {\bf 83}, 235124 (2011).
(arXiv:1103.0472).


\bibitem{Chi}
T. S. Chihara,
{\it An Introduction to Orthogonal Polynomials}
(Gordon and Breach, New York, 1978)


\bibitem{Gt}
W. Gautschi, 
Computational aspects of three-term recurrence relations,
SIAM Review {\bf 9}, 24-82 (1967).


\bibitem{Ht}
A. Hautot,
On the Hill-determinant method,
Phys. Rev. D {\bf 33}, 437-443 (1986).


\bibitem{AMef}
A. Moroz,
Quantum models with spectrum generated by the flows of polynomial zeros,
J. Phys. A: Math. Theor. {\bf 47}(49), 495204 (2014).


\end{thebibliography}
\end{document}